\documentclass[namedreferences]{kluwer}

\input{psfig.sty}
\newcommand{\ket}[1]{\left|#1\right\rangle}
\newcommand{\bra}[1]{\left\langle#1\right|}
\def\Hcontr{H_{\rm ctrl}}
\begin{document}
\begin{article}
\begin{opening}
\title{Dissipation in Josephson qubits}
\subtitle{}

\author{Yuriy \surname{Makhlin}$^{1,2}$}
\author{Gerd \surname{Scn\"on}$^{1,3}$}
\author{Alexander \surname{Shnirman}$^{1}$}

\runningauthor{Yu. Makhlin, G. Sch\"on and A. Shnirman}

\runningtitle{Dissipation in Josephson qubits}

\institute{$^1$\hbox{Institut f\"ur Theoretische Festk\"orperphysik,
Universit\"at Karlsruhe, D-76128 Karlsruhe, Germany}\\
$^2$Landau Institute for Theoretical Physics, Kosygin st. 2, 117940 Moscow,
Russia\\
$^3$Forschungszentrum Karlsruhe, Institut f\"ur
Nanotechnologie, D-76021 Karlsruhe, Germany}

\begin{abstract}
Josephson-junction systems have been studied as potential realizations
of quantum bits. For their operation as qubits it is crucial to
maintain quantum phase coherence for long times. Frequently relaxation and
dephasing effects are described in the frame of the Bloch equations.
Recent experiments demonstrate the importance of $1/f$
noise, or operate at
points where the linear coupling to noise sources is suppressed. This
requires generalizations and extensions of known methods and
results. In this tutorial we present the Hamiltonian for
Josephson qubits in a dissipative environment and review the
derivation of the Bloch equations as well as systematic
generalizations. We discuss $1/f$ noise,
nonlinear coupling to the noise source, and effects of strong pulses on the
dissipative dynamics. The examples illustrate the renormalization of
qubit parameters by the high-frequency noise spectrum
as well as non-exponential decay governed by low-frequency modes.
\end{abstract}
\end{opening}

\section{Introduction}
\label{app:IdealModel}

Quantum-state engineering requires coherent manipulations of suitable
quantum systems.  The needed quantum manipulations can be performed if
we have sufficient control over the fields which couple to the quantum
degrees of freedom, as well as the interactions. The effects of
external noise sources have to be minimized in order to achieve long
phase coherence times. In this tutorial we review the requirements for
suitable physical realizations of qubits -- with emphasis on Josephson
circuits -- and discuss ways to analyze the dissipative effects.

As has been stressed by~\inlinecite{DVD-Curacao}
any physical system, considered as a candidate for quantum
computation, should satisfy the following five criteria:
(i) One needs well-defined two-state quantum systems (qubits).
(ii) One should be able
to prepare the initial state of the qubits with sufficient accuracy.
(iii) A long phase coherence time is needed, sufficient to allow for a
large number (depending on details, say, $\ge 10^4$) of coherent manipulations.
(iv) Control over the qubits' Hamiltonian is
required to perform the necessary unitary transformations.
(v) Finally, a quantum measurement is needed to read out the quantum
information.

The listed requirements may be satisfied by a system of spins -- or
quantum degrees of freedom which under certain conditions
effectively reduce to two-state quantum systems --
which are governed by a Hamiltonian of the form
\begin{equation}
H = \Hcontr(t) + H_{\rm meas}(t) + H_{\rm res} \ .
\end{equation}
The first term, $\Hcontr(t)$, describes the control fields and interactions,
\begin{equation}
\label{Eq:idealH}
\Hcontr(t) = - \frac{1}{2}\sum\limits_{i, a}
                B^i_a(t) \sigma^i_a +\sum\limits_{i\ne j; a,b}
                J^{ij}_{ab}(t) \sigma_a^i \sigma_b^j
                \;.
\end{equation}
Here $\sigma_{a}$ (with $a,b=x,y,z$) are Pauli matrices in the basis of
states $\ket{\uparrow}=\left(1\atop 0\right)$ and
$\ket{\downarrow}=\left(0\atop
1\right)$.
Full control of the unitary quantum  dynamics of individual spins is
achieved if the effective `magnetic field' $\vec B^i(t)$ can be switched
arbitrarily at each
site~$i$. For most purposes control over two
of the field components is sufficient, e.g.,
$H_{\rm ctrl}(t)= - \frac{1}{2} \sum\limits_i  \left[ B^i_z(t) \sigma^i_z +
B^i_x(t) \sigma^i_x \right] $.
In order to perform logic operations, e.g. for quantum
computing, one also needs two-qubit operations. They can be controlled
if the coupling, $ J^{ij}_{ab}(t)$, between
the qubits can be switched. Examples for a suitable coupling are
an Ising $zz$-coupling or a spin-flip $xy$-coupling (see below).

The measurement device and the residual interactions with the
environment are accounted for by extra terms $H_{\rm meas}(t)$
and $H_{\rm res}$, respectively.
Ideally the measurement device should be switchable as well and
be kept in the off-state during manipulations.
The residual interaction $H_{\rm res}$ leads to dephasing and relaxation
processes.  It has to be weak in order to allow for a series of coherent
manipulations.

A typical experiment involves preparation of an initial quantum state,
switching the fields $\vec B^i(t)$ and  coupling energies
$J^{ij}_{ab}(t)$ to effect a specified unitary evolution of the wave function,
and the measurement of the final state.

The initial state can be prepared by keeping the system at low temperatures
in strong enough fields, $B^i_z\gg k_{\rm B}T$, for sufficient time
such that  the residual interaction,
$H_{\rm res}$, relaxes each qubit to its ground state,
$\ket{\uparrow}$.

A single-bit operation on a selected qubit $i$ can be performed, e.g.,
by turning on the field $B^i_x(t)$ for a time span $\tau$.  As a
result the spin state evolves according to the unitary transformation
\begin{equation}
\label{Eq:x-rotation}
U^i_x\left(\alpha\right)= \exp\left(\frac{i B^i_{x} \tau
\sigma^i_x}{2\hbar}\right) =
\left(
\begin{array}{cc} \cos\frac{\alpha}{2} & i\sin\frac{\alpha}{2} \\
 i\sin\frac{\alpha}{2}&\cos\frac{\alpha}{2}
\end{array}
\right)
\, ,
\end{equation}
where $\alpha=B^i_x \tau/\hbar$.  By appropriate choice of the parameters
an $\alpha=\pi$- or an $\pi/2$-rotation can be induced,
producing a spin flip (NOT-operation) or (starting from the ground
state $\ket{\uparrow}$) an equal-weight superposition of spin states,
respectively. Turning on $B^i_z(t)$ for some time $\tau$
produces another  needed single-bit
operation, a phase shift between $\ket{\uparrow}$ and
 $\ket{\downarrow}$, described by $
 U^i_z\left(\beta\right)=   \exp\left(i B^i_z \tau
  \sigma^i_z/2\hbar\right) $
where $\beta=B^i_z \tau/\hbar$. With a sequence of $x$- and
$z$-rotations any unitary transformation of the single-qubit state  can  be
achieved.

A two-bit operation on qubits $i$ and $j$ is induced by turning on
the corresponding coupling $J^{ij}_{ab}(t)$. For instance, for the
$xy$-coupling
$J^{ij} (\sigma_x^i \sigma_x^j+\sigma_y^i \sigma_y^j)$ the result is
described, in the basis $\ket{\uparrow_i
\uparrow_j}$,  $\ket{\uparrow_i \downarrow_j}$, $\ket{\downarrow_i
\uparrow_j}$, $\ket{\downarrow_i \downarrow_j}$,  by the unitary operator
\begin{equation}
\label{2bit_Operation}
U_{\rm 2b}^{ij}\left(\gamma\right)=
\left(
\begin{array}{cccc}
1 & 0 & 0 & 0 \\  0 & \cos\gamma  & i\sin\gamma & 0 \\  0 &
i\sin\gamma & \cos\gamma  & 0 \\ 0 & 0 & 0 & 1
\end{array}
\right)
\,,
\end{equation}
with $\gamma\equiv 2 J^{ij}\tau/\hbar$. For $\gamma=\pi/2$ the
operation leads to a swap of the states $\ket{\uparrow_i\downarrow_j}$ and
$\ket{\downarrow_i\uparrow_j}$ (and multiplication by $i$), while for
$\gamma=\pi/4$ it transforms the state $\ket{\uparrow_i \downarrow_j}$ into  an
entangled state $\frac{1}{\sqrt{2}} \left(\ket{\uparrow_i \downarrow_j} + i
\ket{\downarrow_i \uparrow_j} \right)$.

Instead of the sudden switching of
$B_{z,x}^i(t)$ and $J^{ij}(t)$, discussed above for illustration,
one can use other techniques to
implement single- or two-bit operations.  For instance,
ac resonance signals can  induce Rabi
oscillations between different states of a qubit or qubit pairs.
Both switching and ac-techniques have been applied for Josephson
qubits, e.g., by~\inlinecite{Nakamura_Nature} and
\inlinecite{Saclay_Manipulation_Science},~\inlinecite{Delft_Rabi}, respectively.

The coupling to the environment, described by $H_{\rm res}$,
leads to dephasing and relaxation processes. In this tutorial we will
first derive in Section~\ref{Sec:SB}
the proper form of $H_{\rm res}$ for the case where a
Josephson charge qubit is coupled to an electromagnetic environment
characterized by an arbitrary impedance.
In Section~\ref{Sec:Diss} we present a systematic perturbative approach, derive
the Bloch equations and discuss their validity range.
In particular, we find expressions for the relaxation and dephasing
rates. In the  following Section~\ref{Sec:Appl} we use the Bloch equations
to study several problems of interest. Section~\ref{Sec:beyond} deals with
extensions beyond the Bloch-Redfield description.

\section{Dissipation in Josephson circuits}
\label{Sec:SB}

\begin{figure}
\centerline{\hbox{\psfig{figure=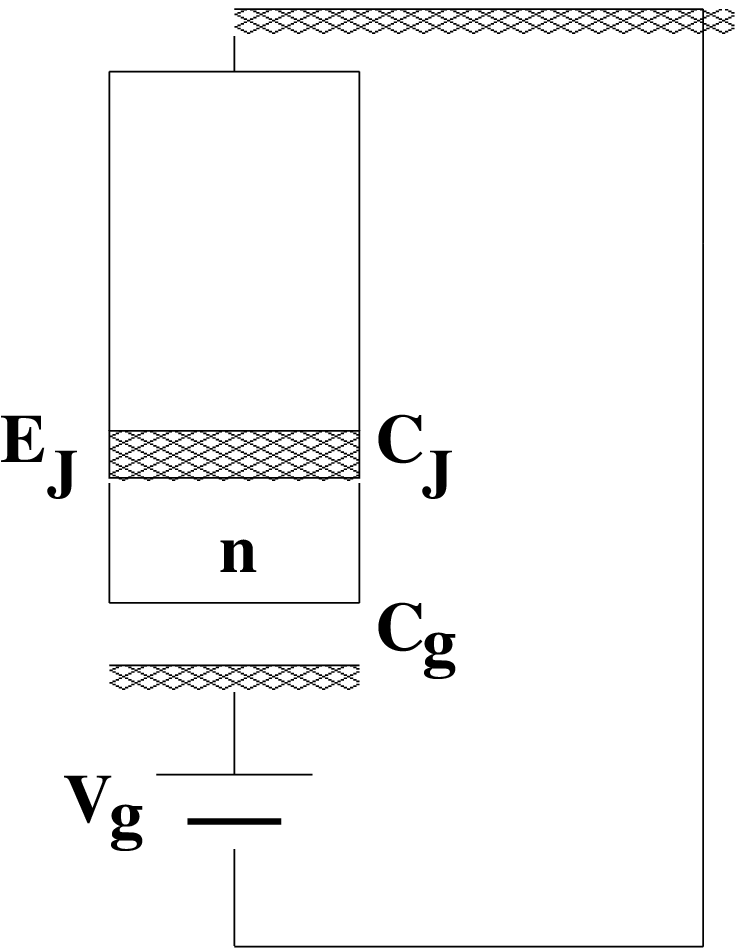,%
height=5cm}\hskip 1cm\psfig{figure=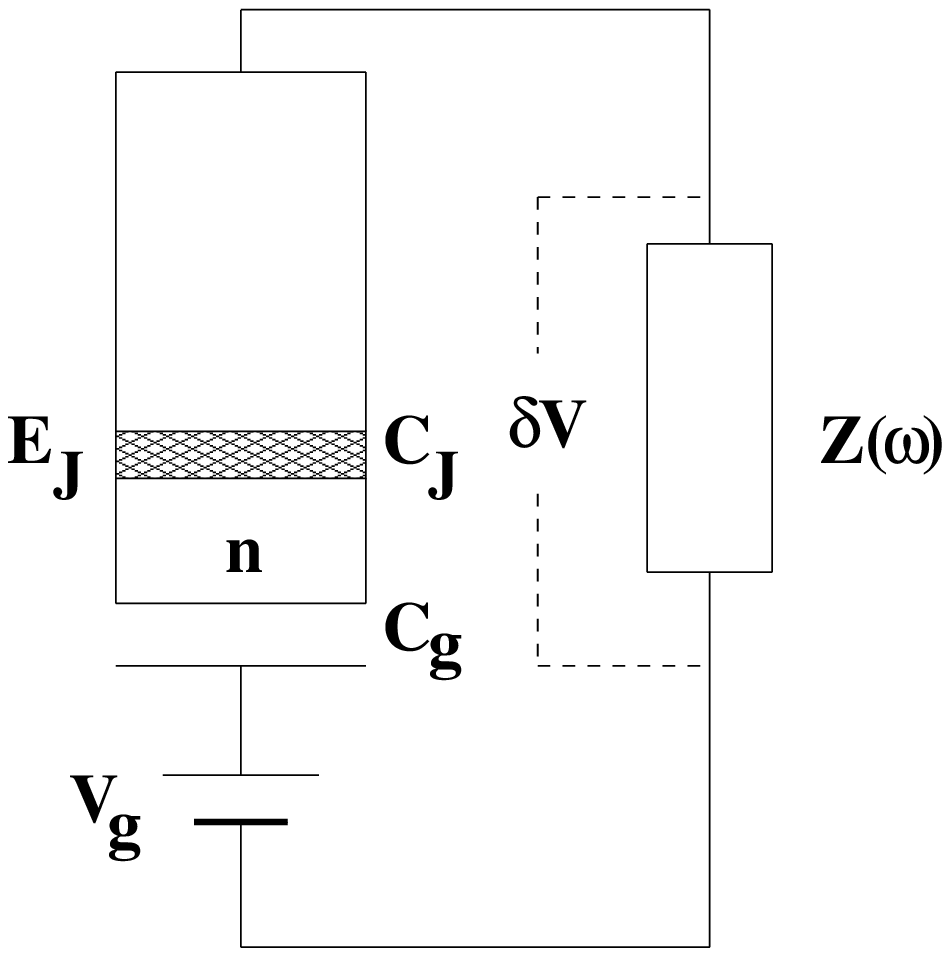,%
height=5cm}}}
\caption[]{a) A Josephson charge qubit in its simplest design
formed by a superconducting single-charge box. b) A qubit with
the electromagnetic environment represented by the impedance $Z(\omega)$.}
\label{Fig:Charge_Qubit}
\end{figure}

\subsection{Josephson qubits, the Hamiltonian and the dissipation}
Fig.~\ref{Fig:Charge_Qubit} shows an example of a `Josephson charge
qubit' built from superconducting tunnel junctions. All Josephson qubits
have in common that they are nanometer-size
electronic elements embedded into and manipulated by electrical
circuits. The electromagnetic environment, which can not be avoided if
we want to control the qubit, leads also to dissipation.
This environment plays a crucial role in many contexts, e.g., the physics of the
Coulomb blockade in tunnel junctions, for which the appropriate
theory is reviewed by~\inlinecite{Ingold_Nazarov}. In the spirit
of this analysis we will analyze
the effects of the environment on Josephson qubits.

We first consider a superconducting single-charge box shown in
Fig.~\ref{Fig:Charge_Qubit}a in the absence of dissipation.
We further choose parameters such that
single-electron tunneling is suppressed and only Cooper-pair charges
tunnel across the junction. This situation is realized if the superconducting
gap $\Delta$ is larger than the charging energy scale $E_{\rm C}$
(defined below) and the temperature $k_{\rm B}T$. In this case
superconducting charge box is described by the
Hamiltonian
\begin{equation}
\label{Eq:Box_Hamiltonian_n}
H_{\rm box}=
4E_{\rm C}\left(n-\frac{Q_{\rm g}}{2e}\right)^2 - E_{\rm J} \cos\theta
\ .
\end{equation}
Here n is the operator counting the number of excess Cooper pairs on
the island relative to a neutral reference state. It is conjugate to
$\theta$, the phase difference of the superconducting order
parameter across the junction. Charge quantization requires that
physical quantities are $2\pi$-periodic
functions of $\theta$, for instance $e^{i\theta}$, with $e^{i\theta}|n\rangle =
|n+1\rangle$. The `gate charge'
$Q_{\rm g}\equiv C_{\rm g}V_{\rm g}$
is controlled by the gate voltage, and $E_{\rm C}\equiv
e^2/2(C_{\rm J}+C_{\rm g})$, which depends on the junction and gate
capacitances,  is the single-electron charging
energy scale. Interactions with charges in the substrate induced by
stray capacitances, leads to additional contributions to the gate
charge.

Next we account for the dissipation due to electromagnetic
fluctuations. They can be modeled by an
effective impedance $Z(\omega)$, placed in series with the voltage source
(see Fig.~\ref{Fig:Charge_Qubit}b) and producing a fluctuating
voltage. The sum
$V_{\rm g} + \delta V(t)$ couples to the charge in the circuit.
For an impedance with open leads the Johnson-Nyquist
expression relates the voltage fluctuations between its terminals to
$Z(\omega)$. In the present case the impedance is embedded in
a circuit, which further modifies  the spectrum of  voltage
fluctuations. Neglecting for a moment
the Josephson tunneling (i.e., treating the Josephson junction as a capacitor)
we obtain from the fluctuation-dissipation theorem (FDT)
\begin{equation}
\label{Eq:S_V}
S_V(\omega) \equiv \frac{1}{2} \int dt \, e^{i\omega t}
\langle \left\{\delta V(t), \delta V(0)\right\}\rangle=
{\rm Re}Z_{\rm t}(\omega)\, \hbar \omega
\coth \frac{\hbar \omega}{2k_{\rm B}T}
\ .
\end{equation}
Here
$Z_{\rm t}(\omega) \equiv \left[-i\omega C_{\rm t} +
Z^{-1}(\omega)\right]^{-1}$ and
$C_{\rm t}^{-1} = C_{\rm J}^{-1} + C_{\rm g}^{-1}$
are the total impedance and capacitance as seen
from the terminals of $Z(\omega)$, respectively.

For many purposes it is sufficient to characterize the environment by
its noise  spectrum and the corresponding response function. In
general one should take into account that the environment
itself is also a quantum system with many degrees
of freedom. As argued before~\cite{Caldeira-Leggett_AnnPhys83}, in a generic
setting one can think of a linear environment as of a large set of harmonic
oscillators, each of which interacts weakly with the system of
interest. I.e., $H_{\rm bath}\equiv \sum_a \left[
\frac{p_a^2}{2m_a} +\frac{m_a\omega_a^2}{2}x_a^2\right]$, and the qubit is
coupled to the fluctuation variable $\delta V = \sum_a \lambda_a x_a$.
The Johnson-Nyquist relation (\ref{Eq:S_V}) is reproduced if
one chooses the bath `spectral density' as $J_V(\omega) \equiv
{\pi\over 2}\sum_a \frac{\lambda_a^2}{m_a\omega_a} \delta(\omega-\omega_a) =
\omega\, {\rm Re}Z_{\rm t}(\omega)$.  The Hamiltonian of the
whole system then reads\footnote{Here we `postulate' the model based on
experience with
dissipation in quantum systems. For the reader who is less
familiar with these ideas we present in the appendix a more detailed
derivation of the Hamiltonian for a particular model of the
electromagnetic environment, namely a resistor modeled as a
transmission line.}
\begin{equation}
\label{Eq:Hamiltonian_CL}
H = H_{\rm box}  -
2en {C_{\rm t} \over C_{\rm J}}\, \delta V
+H_{\rm bath}+
\frac{1}{2\tilde C_{\rm t}}
\left(2en{C_{\rm t} \over C_{\rm J}}\right)^2
\ .
\end{equation}
The last `counter-term' is introduced to cancel the renormalization
of the qubit's charging energy by the environment.
The capacitance $\tilde C_{\rm t}$ is defined by $\int_0^\infty d\omega\,
J_V(\omega)/\omega \equiv \pi/(2\tilde C_{\rm t})$. Because of the analytic
properties of response functions it also equals  $\tilde C_{\rm t}=
C_{\rm t} + \lim_{\omega\to\infty}i/(\omega Z(\omega))$.
For a pure resistor $Z(\omega)=R$ it reduces
to $C_{\rm t}$. The last three terms in (\ref{Eq:Hamiltonian_CL}) can
be lumped as
$$
H = H_{\rm box}  +
\sum_a \left[
\frac{p_a^2}{2m_a} +\frac{m_a\omega_a^2}{2}
\left( x_a - \frac{\lambda_a}{m_a\omega_a^2} \, 2en \frac{C_{\rm t}}{C_{\rm J}}
\right)^2\right] \, .
$$

\subsection{Reduction to a two-level system}

If we bias the superconducting charge box near a degeneracy point of
the charging energy, e.g. at a voltage close to
$e/C_{\rm t}$, and choose all other parameters (temperature,
frequency and strength of control pulses) appropriate,
only two charge states play a role, $\ket{n=0}$ and $\ket{n=1}$, which
we denote as $\ket{\uparrow}$ and $\ket{\downarrow}$, respectively.
The leakage to higher charge states can be kept at a low
level~\cite{Saro_Leakage}.
In spin notation, the number operator
becomes $n=\frac{1}{2}(1-\sigma_z)$, while
$\cos\theta = \frac{1}{2}\sigma_x$, and the effective spin Hamiltonian reads
\begin{equation}
\label{Eq:Spin_Boson_Hamiltonian}
H = -\frac{1}{2}B_z(V_{\rm g})\;\sigma_z-\frac{1}{2}B_x\;\sigma_x
-\frac{1}{2}X\;
\sigma_z + H_{\rm bath}
\ ,
\end{equation}
where $X= -\frac{C_{\rm t}}{C_{\rm J}}\, 2e\delta V\,$, $B_x = E_{\rm J}$, and
\begin{equation}
\label{Eq:B_z_Q_g}
B_z = \frac{2e}{C_{\rm J}+C_{\rm g}}(e-C_{\rm g}V_{\rm g})=
4 E_{\rm C}\left(1-\frac{Q_{\rm g}}{e}\right)
\ .
\end{equation}
The spectral density of the fluctuating field $X$ is related to the
power spectrum of voltage fluctuations by
\begin{equation}
\label{Eq:S_X}
S_X(\omega) = \frac{1}{2} \left( \left\langle X^2_\omega\right\rangle
+ \left\langle X^2_{-\omega}\right\rangle\right)
= \left(2e\,\frac{C_{\rm
t}}{C_{\rm J}}\right)^2\,S_V(\omega) \ .
\end{equation}
Here we defined
$\left\langle X^2_\omega \right\rangle \equiv \int dt\, e^{i\omega t}
\left\langle X(t)X(0) \right\rangle$.
For the calculation of many properties of the two-level system the
knowledge of  the symmetrized noise correlator,
$S_X$, is sufficient. In general, however, the anti-symmetric
part $A_X(\omega) \equiv \frac{1}{2}\,\langle \left[X(t),
X(0)\right]\rangle_\omega$ is needed as well.
It is related to the response function of the bath
$\chi(t)=(i/\hbar)\theta(t)\langle[X(t), X(0)]\rangle$,
which describes the reaction of the bath to a perturbing force.
Namely, if a perturbation $H_{\rm p}=-fX$ is added to the Hamiltonian
the bath responds as $\langle X \rangle_{\omega}=\chi(\omega) f(\omega)$.
The function $A_X(\omega)$ is related to the imaginary part of the
response function, $A_X(\omega)=\hbar\chi''(\omega)$,
and in equilibrium the FDT fixes
$S_X(\omega) = A_X(\omega)\coth(\hbar \omega/2k_{\rm B}T)$.

To simplify the comparison with the literature on the Caldeira-Leggett model
we include the prefactor in (\ref{Eq:S_X}) into the
definition of the spectral density,
$J_X(\omega) \equiv (2e C_{\rm t}/C_{\rm J})^2 J_V(\omega) = \chi''(\omega)$.
The generic low-frequency behavior is a power-law, $J_X(\omega) =
2\pi\hbar\alpha \omega_0^{1-s}\omega^s$, where
$\omega_0$ is a frequency scale and $\alpha$ the dimensionless
strength of the dissipation. Of particular interest is the Ohmic case ($s=1$),
obtained if $Z(\omega)=R$ and, hence, ${\rm Re}\, Z_{\rm t}(\omega)
\approx R$, for $\omega \ll (RC_{\rm t})^{-1}$. In this case we have $J_X =
2\pi\hbar\alpha\omega\, $ and
\begin{equation}
S_X(\omega) = 2\pi\hbar^2\alpha\omega\coth\frac{\hbar\omega}{2k_{\rm B}T}\,,
\end{equation}
with
\begin{equation}
\label{Eq:alpha}
\alpha = \frac{R}{R_{\rm Q}}\,\left(\frac{C_{\rm t}}{C_{\rm
J}}\right)^2 \ ,
\end{equation}
and $R_{\rm Q}\equiv (2e)^2/h$ is the (superconducting)
resistance quantum.


\section{Dissipative dynamics}\label{Sec:Diss}
\subsection{Bloch equations}

The dissipative dynamics of spins has been the subject of extensive
research in the context of the nuclear magnetic resonance (NMR).
One of the main tools in this field are the Bloch equations.
These kinetic equations were first formulated, on phenomenological grounds,
by~\inlinecite{Bloch_Initial} for the case of nuclei in a
magnetic field $\vec B = \vec B_{\parallel} + \delta \vec B_{\perp}(t)$,
which is the sum of a strong static field $\vec
B_{\parallel}$ in the $z$-direction, and a weak, time-dependent, transverse
perturbation
$\delta \vec B_{\perp}(t)$. The latter may be chosen to oscillate in
resonance with the Larmor frequency $B_{\parallel}/\hbar$ to induce spin
flips. The Bloch
equations describe the dynamics of the magnetization in the `anisotropic
$\tau$-approximation':
\begin{equation}
\label{Eq:Bloch_Equations}
\frac{d}{dt} \vec M = -\vec B \times \vec M - \frac{1}{T_1}
(M_z - M_0) \vec z
-\frac{1}{T_2}(M_x \vec x + M_y \vec y)
\ .
\end{equation}
The two relaxation times,
$T_1$ and $T_2$, characterize the relaxation of the longitudinal
component of the magnetization $M_z$
to $M_0$, which is the equilibrium  magnetization in the static
field $\vec B_\parallel$, and the transverse components
($M_x$, $M_y$) to zero, respectively.
 Eq.~(\ref{Eq:Bloch_Equations}) describes
the evolution of a two-level system in terms of its `magnetization'
$\vec M\propto \left\langle \vec\sigma \right\rangle$, and at the same
time it describes the time evolution of the components of its density matrix,
related to the magnetization by $\left\langle \sigma_z \right\rangle
=\rho_{00}-\rho_{11}$ and
$\left\langle \sigma_x+i\sigma_y \right\rangle =\rho_{10}$.
Here $0$ and $1$ denote the
ground state, $\ket{\uparrow_z}$, and excited state,
$\ket{\downarrow_z}$, in the field $B_{\parallel}$, respectively.

Using the normalization, $\mathop{\rm tr}\hat\rho=\rho_{00}+\rho_{11}=1$, one
can rewrite
Eq.~(\ref{Eq:Bloch_Equations}) as equations of motion for the density matrix:
\begin{eqnarray}
\label{Eq:Spin_Master_Equation}
\dot \rho_{00} &=& -\Gamma_{\uparrow} \rho_{00} + \Gamma_{\downarrow} \rho_{11}
\ ,\nonumber\\
\dot \rho_{11} &=& \Gamma_{\uparrow} \rho_{00} - \Gamma_{\downarrow} \rho_{11}
\ ,\nonumber\\
\dot \rho_{01} &=& -i B_{\parallel} \rho_{01} - \frac{1}{T_2} \rho_{01}
\ ,
\end{eqnarray}
where the excitation rate $\Gamma_\uparrow$ and
relaxation rate $\Gamma_\downarrow$ are related to $T_1$ and the
equilibrium value $\langle\sigma_z\rangle_0$ by
$1/T_1=\Gamma_{\downarrow}+\Gamma_{\uparrow}$ and
$\langle\sigma_z\rangle_0 = T_1 (\Gamma_{\downarrow}-\Gamma_{\uparrow})$.

A series of papers were devoted to the derivation and
generalization of the Bloch
equations~\cite{Wangsness_Bloch,Bloch_Derivation,Redfield_Derivation}.
Below we will illustrate the derivation from the model
(\ref{Eq:Spin_Boson_Hamiltonian}) in several limits of
the qubit's dynamics. We can anticipate two different regimes: The {\it
Hamiltonian-dominated} regime where the dissipative effects are slow
compared to the Larmor precession. In this case  it is convenient to describe
the dynamics in the eigenbasis
of the spin's Hamiltonian. The other, {\it dissipation-dominated} regime,
arises, when the total magnetic field is weak. Then
the dissipation dominates, and the preferred eigenbasis is that of the
dissipative part of the Hamiltonian ($\propto \sigma_z$).

\subsection{Golden Rule and the Bloch equations}
\label{Sec:GR}

In the eigenbasis of the spin (qubit) the Hamiltonian
(\ref{Eq:Spin_Boson_Hamiltonian}) reads
\begin{equation}
\label{Eq:Spin_Boson_Eigen_Basis}
H= -\frac{1}{2}\Delta E\;\sigma_z - \frac{1}{2}
X\,\left(\cos\eta\; \sigma_z - \sin\eta\; \sigma_x \right)
+ H_{\rm bath}
\ ,
\end{equation}
where $\Delta E \equiv \sqrt{B_z^2+B_x^2} \,$ and $\, \tan\eta = B_x/B_z$.
We denote the ground and excited states of the free qubit by
$\ket{0}$ and $\ket{1}$, respectively.
The coupling to the bath is a sum of a transverse
($\propto \sin\eta$) and a longitudinal ($\propto \cos\eta$) term.
Only the transverse part can cause spin flips.
In the weak-noise limit we consider $X$  as a perturbation and
apply Fermi's golden rule to obtain the relaxation rate,
$\Gamma_\downarrow = \Gamma_{\ket{1}\to \ket{0}}$,
and  excitation rate,
$\Gamma_\uparrow = \Gamma_{\ket{0}\to \ket{1}}$.
E.g., for $\Gamma_\downarrow$ we obtain
\begin{eqnarray}
\label{Eq:Gamma_down}
\Gamma_{\downarrow} &=&
\frac{2\pi}{\hbar}\;\frac{\sin^2\eta}{4}\;
\sum_{i,f}\rho_i\; |\bra{i} X \ket{f}|^2\;
\delta(E_i + \Delta E - E_f)
\nonumber \\
&=&
\frac{2\pi}{\hbar}\;\frac{\sin^2\eta}{4}\;
\sum_{i,f}\rho_i\;\bra{i} X \ket{f}\bra{f}X\ket{i}\;
\frac{1}{2\pi\hbar}\int dt\; e^{i\frac{t}{\hbar}\;(E_i + \Delta E - E_f)}
\nonumber\\
&=&
\frac{\sin^2\eta}{4\hbar^2}\;\int dt\;
\sum_{i}\rho_i\;\bra{i} X(t) X \ket{i}\;
e^{i\frac{t}{\hbar}\;\Delta E}
\nonumber\\
&=&
\frac{\sin^2\eta}{4\hbar^2}\;
\langle X^2_{\omega = \Delta E/\hbar}\rangle
\ .
\end{eqnarray}
Here $\ket{i}$ and $\ket{f}$ are the initial and the final states
of the bath and $\rho_i$ is the probability for the bath to be
initially in the state $\ket{i}$. Similarly, we obtain
\begin{equation}
\label{Eq:Gamma_up}
\Gamma_{\uparrow} = \frac{\sin^2\eta}{4\hbar^2}\;
\langle X^2_{\omega = -\Delta E/\hbar}\rangle
\ .
\end{equation}
For the relaxation time we thus find
\begin{equation}
\label{Eq:T1} \frac{1}{T_1}
=\Gamma_{\downarrow}+\Gamma_{\uparrow}=\frac{\sin^2\eta}{2\hbar^2}\;
S_X(\omega = \Delta E/\hbar) \ ,
\end{equation}
and for the equilibrium magnetization
\begin{equation}
\langle\sigma_z\rangle_0 =
\frac{\Gamma_{\downarrow}-\Gamma_{\uparrow}}{\Gamma_{\downarrow}
+\Gamma_{\uparrow}}= \frac{A_X(\omega=\Delta
E/\hbar)}{S_X(\omega=\Delta E/\hbar)}=\tanh\frac{\Delta E} {2k_{\rm B}T}
\ .
\end{equation}

The golden rule can also be used to evaluate the dephasing time.
Here we skip the
derivation since it will be performed in Section~\ref{Sec:T1T2} in
the framework of the Bloch-Redfield approximation. The result is
\begin{equation}
T_2^{-1}=\frac{1}{2}T_1^{-1} + \frac{1}{2\hbar^2}
\cos^2\eta\, S_X(\omega=0) \,.
\label{T2GR}
\end{equation}
Note that the first term $\propto S(\omega=\Delta E/\hbar)$ in
Eq.~(\ref{T2GR}) is the result of the transverse noise, which involves
transitions  between the qubit's states with the energy change of
$\Delta E$. The second term
$\propto S(\omega=0)$ is associated with the longitudinal noise, which
does not flip the spin and therefore involves only transitions in the bath
without energy transfer.
It produces a random contribution to energy
differences, and hence to the accumulated phase difference.
This contribution to the dephasing rate is sometimes called the
``pure-dephasing'' rate, $1/T_2^*$, so that $T_2^{-1} = \frac{1}{2}\,T_1^{-1} +
(T_2^*)^{-1}$.

\subsection{Diagrammatic technique}
\label{Sec:Diag}

\begin{figure}
\centerline{\hbox{\psfig{figure=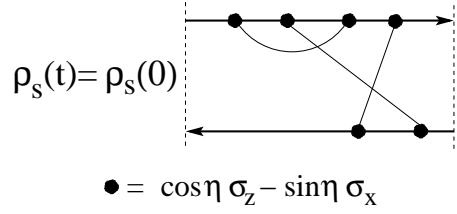,width=0.5\columnwidth}}}
\caption[]{%
Diagrammatic representation of the propagator $\Pi$.}
\label{Fig:Diagrams_Pi}
\end{figure}
We proceed with a systematic perturbative analysis
of the time evolution of the qubit's density matrix based on the Keldysh
diagrammatic  technique. We derive the Bloch equations and the
corresponding dissipative times  within the so-called Bloch-Redfield
approximation, which may be thought of as a generalized golden rule.
The systematic extensions allow us to analyze effects of
higher orders in the noise and of long-time noise correlations. We
will also determine the limits of applicability of the Bloch equations.

First we
briefly sketch the derivation of the master equation for the evolution of the
density matrix, based on the formalism developed by~\inlinecite{Schoeller_PRB}.
With the interaction Hamiltonian
\begin{equation}
H_{\rm int} = -\frac{1}{2}\,X
\left(\cos\eta\; \sigma_z - \sin\eta\; \sigma_x \right)
\label{Eq:Hint}
\end{equation}
the time evolution of the density matrix in the interaction
representation  follows from
\begin{equation}
\label{Eq:Rho_Time_Evolution}
\rho(t) =  T e^{-\frac{i}{\hbar}\int\limits_{t_0}^{t} H_{\rm int}(t') dt'}
\rho(t_0)\;
\tilde T e^{\frac{i}{\hbar}\int\limits_{t_0}^{t} H_{\rm int}(t') dt'}
\ .
\end{equation}

Assuming that initially, at $t=t_0$, the density matrix is factorized,
$\hat\rho(t_0) = \rho_s(t_0) \otimes \rho_{\rm bath}$, we trace
Eq.~(\ref{Eq:Rho_Time_Evolution}) over the bath degrees of freedom
and obtain a perturbative expansion for the propagator $\Pi$ of
the spin's density matrix $\rho_{\rm s}(t)=\Pi(t,t_0)\rho_{\rm s}(t_0)$.
On one hand, the  assumption of a factorized initial condition
allows us to introduce  the characteristics
of the bath, e.g. its temperature. On the other hand, it may introduce
artifacts. In general we find for $\rho_{\rm
s}$ a non-Markovian (nonlocal in time) evolution equation
such that the behavior at a time $t$ involves knowledge of $\rho_{\rm s}$
at earlier times. However, often the kernel of this integral equation has a
limited range $\tau_{\rm c}$ and does not depend on the initial spin
state. Then the simple assumption about the initial density matrix
should not influence the long-time evolution.

\begin{figure}
\centerline{\hbox{\psfig{figure=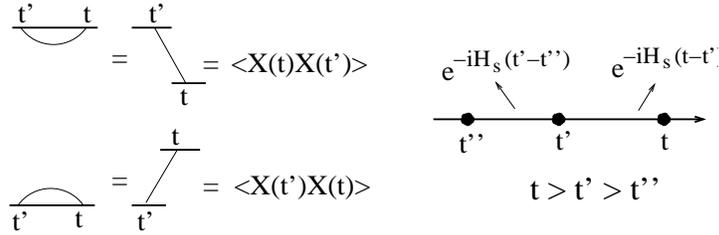,width=0.8\columnwidth}}}
\caption[]{%
Basic elements of the diagrammatic technique.}
\label{Fig:Diagrams_Lines}
\end{figure}

By tracing out the bath degrees of freedom in
Eq.~(\ref{Eq:Rho_Time_Evolution}) and expanding in $H_{\rm int}$,
one obtains a perturbative series. A typical contribution to $\Pi$ is shown
in Fig.~\ref{Fig:Diagrams_Pi}. The upper line contains vertices (the terms
$H_{\rm int}$) from the first, time-ordered exponent in
Eq.~(\ref{Eq:Rho_Time_Evolution}), and the lower one contains vertices
from the other,
inversely time-ordered exponent. As one can see from Eq.~(\ref{Eq:Hint}), each
vertex in the diagrams contains one bath operator $X$.
The emerging averages of their products reduce, due to Wick's theorem,
in a standard way to pairwise averages, which are  shown by the thin
lines in the figure. In Fig.~\ref{Fig:Diagrams_Lines}a
the diagrammatic rules for the thin lines are sketched. The horizontal solid
lines (see Fig.~\ref{Fig:Diagrams_Lines}b) describe explicitly
the evolution of the spin degree of freedom, each
line
corresponding to (a $2\times 2$ operator) $\exp[\pm i H_{\rm s}(t-t')]$
with different signs for the upper and lower lines.

\begin{figure}
\centerline{\hbox{\psfig{figure=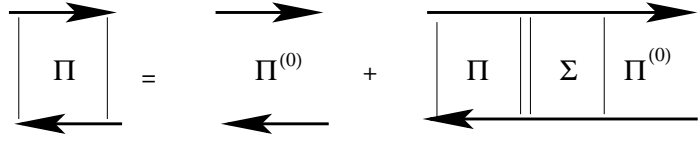,width=0.8\columnwidth}}}
\caption[]{%
The Dyson equation for the propagator $\Pi$.}
\label{Fig:Dyson}
\end{figure}

\begin{figure}
\centerline{\hbox{\psfig{figure=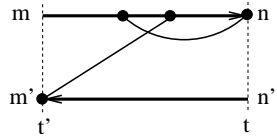,width=0.3\columnwidth}}}
\caption[]{%
A (second-order) contribution to $\Sigma_{nn'\leftarrow mm'}(t-t')$.}
\label{Fig:Sigma}
\end{figure}

The propagator $\Pi$ satisfies the Dyson equation shown in
Fig.~\ref{Fig:Dyson}.
Taking the time derivative one arrives at the
master (kinetic) equation
\begin{eqnarray}
\frac{d}{dt}\rho_{\rm s}(t)&=&\frac{i}{\hbar} \left[\rho_{\rm
s}(t),H_{\rm s}\right]+ \int\limits_{0}^t dt'\;
\Sigma(t-t')\;\rho_{\rm s}(t') \nonumber\\
&=&L_{\rm s} \rho_{\rm s}(t) +
\int\limits_{0}^t dt'\; \Sigma(t-t')\;\rho_{\rm s}(t')
\ .
\label{Eq:Dyson_Equation}
\end{eqnarray}
We introduced the Liouville operator $L_{\rm s} \rho_{\rm s} \equiv
(i/\hbar)\,\left[ \rho_{\rm s}, H_{\rm s}\right]$.
The self-energy $\Sigma$ is defined as the sum of all irreducible
diagrams, i.e., diagrams that can not be divided by cutting the upper and
lower solid lines at the same time (see an
example in Fig.~\ref{Fig:Sigma}). It is an operator in the space of
density matrices, i.e., a tensor of rank
four. We denote its components by $\Sigma_{nn' \leftarrow mm'}$, implying the
following action on density matrices:
$\displaystyle (\Sigma\rho_{\rm s})_{nn'}= \sum_{mm'}
\Sigma_{nn' \leftarrow mm'}\, (\rho_{\rm s})_{mm'}$.


\subsection{Bloch-Redfield and the rotating-wave approximations}
\label{Sec:Bloch-Redfield}

Under certain conditions Eq.~(\ref{Eq:Dyson_Equation}) may be approximated by
a simpler equation which is local in time. Indeed,  the
self-energy (which in lowest order is proportional to the
noise correlator) in the last term on
the right hand side decays on time differences $t-t'$ of the order of
$\tau_{\rm c}$. If this time is much shorter than the
dissipative dephasing and relaxation times we may replace the density
matrix in the integral from $t'$ to $t$ by the leading term
\begin{equation}
\label{Eq:Short_Time_Markov}
\rho_{\rm s}(t') \approx e^{-L_{\rm s}(t-t')}\;\rho_{\rm s}(t)
\ ,
\end{equation}
and  obtain a Markovian evolution equation
\begin{equation}
\frac{d}{dt}\rho_{\rm s}(t)=L_{\rm s}\, \rho_{\rm s}(t) +
\hat \Gamma\,\rho_{\rm s}(t) \, .
\label{Eq:Redfield}
\end{equation}
It involves the `Bloch-Redfield tensor'
\begin{equation}
\label{Eq:Bloch_Redfield_Tensor}
\hat \Gamma \equiv \int\limits_{-\infty}^\infty dt\; \hat \Sigma(t)\;
e^{-L_{\rm s}t}\, .
\end{equation}
To simplify the Fourier analysis here and in the following, the definition
of the self-energy is extended to negative times with  $\hat \Sigma(t<0) = 0$.
 In the
eigenbasis of $H_{\rm s}$ the Liouvillian $L_{\rm s}$ is diagonal, $[L_{\rm
s}]_{mm'\leftarrow mm'} = i(E_m-E_{m'})\hbar$, and
the components of the Bloch-Redfield tensor are related to the self-energy:
$\Gamma_{nn'\leftarrow mm'}=
\Sigma_{nn'\leftarrow mm'}(\hbar\omega = E_m-E_{m'})$.

To verify the validity of the Bloch-Redfield approximation
one should check whether the dissipative times $T_1$, $T_2$
in the resulting Bloch equations are indeed much longer than the
correlation time $\tau_{\rm c}$ of the integrand in
Eq.~(\ref{Eq:Bloch_Redfield_Tensor})
[more precisely, whether the integral is dominated by times
shorter than $T_{1}$ and $T_{2}$].~\footnote{Note  that
due to differences in the oscillatory factors different
components of the matrix integral in
Eq.~(\ref{Eq:Bloch_Redfield_Tensor}) can be dominated by different
time scales.  Within the RWA the influence of dissipative terms
on the $nn'$-component of the density matrix is dominated by the behavior of
the
bath spectral power close to $\hbar\omega = E_n-E_{n'}$. One finds that the
validity of the Bloch-Redfield approximation requires that the spectrum
does not change much on the
scale given by the respective component of $\hat \Gamma$.}

In certain cases the equations of motion can be simplified further
by employing the rotating-wave approximation (RWA). It is based on the
following considerations: if the dissipation is weak and the dynamics of the
components of the density matrix close to the unperturbed dynamics, the
spectral weight of $\rho_{nn'}(t)$ is located in the vicinity of
$\omega=(E_n-E_{n'})/\hbar\equiv\omega_{nn'}^0$. The rhs of the $nn'$-component
of the equation of motion (\ref{Eq:Redfield}) contains contributions from all
components $mm'$ of the density matrix. However, if $\omega_{mm'}^0 \neq
\omega_{nn'}^0$ this contribution averages out fast (on the time scale
$\hbar/(\omega_{mm'}^0 - \omega_{nn'}^0)$) and can be neglected.
I.e., one truncates the terms
$\Gamma_{nn'\leftarrow mm'} \rho_{mm'}$ on the rhs of the equation of motion:
 for the diagonal elements  $\rho_{nn}$ one
keeps only the  diagonal entries, while for the off-diagonal elements,
e.g.  $\rho_{01}$,  one
keeps only the contribution from the same entry, e.g. the component $\Gamma_{01
\leftarrow 01}$.
The latter are, in general, complex numbers, with a
real part describing dephasing and an imaginary part corresponding
to a renormalization of the energy.

If the system is subject to a strong pulse,
Eq.~(\ref{Eq:Redfield}) may not be sufficient. Typically the
initial evolution for a certain period of time $\sim\tau_{\rm c}$
needs to be described by other methods. However, the following time
evolution is governed by (\ref{Eq:Redfield}).
We will discuss examples in Section~\ref{Sec:Pulse}.
Another situation, where Eq.~(\ref{Eq:Redfield}) may be not
sufficient is the case where the noise shows long-range
correlations (cf.~the discussion of the 1/f noise in Section~\ref{Sec:1/f}).


\subsection{$T_1$ and $T_2$}\label{Sec:T1T2}

\begin{figure}
\centerline{\hbox{\psfig{figure=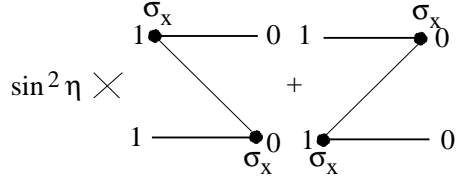,width=0.5\columnwidth}}}
\caption[]{%
Leading-order contributions to $\Gamma_{00 \leftarrow 11}$.}
\label{Fig:Gamma_11to00}
\end{figure}

As an example and demonstration of the diagrammatic expansion
we rederive the expressions for $T_1$ and $T_2$,
obtained before by golden rule arguments in Section~\ref{Sec:GR}.
The lowest-order contributions to the rate $\Gamma_{00 \leftarrow 11}
= \Gamma_{\downarrow}$ are shown in Fig.~\ref{Fig:Gamma_11to00}.
We obtain
\begin{equation}
\label{Eq:Sigma_2211} \Sigma_{00 \leftarrow 11}(t-t') =
\frac{\sin^2\eta}{4\hbar^2}\,
\left[
\langle X(t)X(t')\rangle e^{\frac{i}{\hbar}\Delta E(t-t')}
+c.c.
\right] \ ,
\end{equation}
and thus, from Eq.~(\ref{Eq:Bloch_Redfield_Tensor})
and the fact that ${(L_{\rm s})}_{11 \leftarrow 11}=0$, we get
\begin{eqnarray}
&&\Gamma_{\downarrow}=\Gamma_{00 \leftarrow 11} =
\Sigma_{00 \leftarrow 11}(\omega=0)=
\nonumber\\
&&{\rm Re}\left[-\frac{i\sin^2\eta}{2\hbar^2} \int
\frac{d\omega'}{2\pi}\;\frac{\langle X^2_{\omega'}\rangle}{\omega'-(\Delta
E/\hbar)-i0}\right]=
\nonumber \\
&&\frac{\sin^2\eta}{4\hbar^2}\;
\langle X^2_{\omega= \Delta E/\hbar} \rangle
\ .
\end{eqnarray}
In the same way one can obtain the excitation rate $\Gamma_{11 \leftarrow
00}\equiv\Gamma_\uparrow$, and hence $T_1$,
as well as the rates
$\Gamma_{00 \leftarrow 00}=-\Gamma_{11 \leftarrow 00}$ and
$\Gamma_{11 \leftarrow 11}=-\Gamma_{00 \leftarrow 11}$.

\begin{figure}
\centerline{\hbox{\psfig{figure=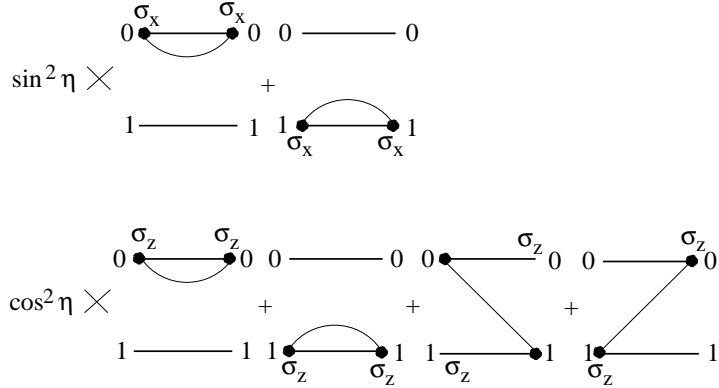,width=0.8\columnwidth}}}
\caption[]{%
Leading-order contributions to  $\Gamma_{01 \leftarrow 01}$.}
\label{Fig:Gamma_01to01}
\end{figure}

As for the dephasing time $T_2$, the relevant lowest-order
contributions to $\Gamma_{01 \leftarrow 01}$ are shown in
Fig.~\ref{Fig:Gamma_01to01}. The upper row of the diagrams shows the
contribution of the transverse fluctuations $\Sigma_{01 \leftarrow 01}^{\perp}$
and the lower row shows that of the longitudinal fluctuations $\Sigma_{01
\leftarrow 01}^{\parallel}$, so that $\Sigma_{01 \leftarrow 01}=
\Sigma_{01 \leftarrow 01}^{\perp}+\Sigma_{01 \leftarrow 01}^{\parallel}$.

The transverse and longitudinal contributions give:
\begin{equation}
\label{Eq:Transverse_Sigma} \Sigma_{01 \leftarrow 01}^{\perp}(t-t') =
-\frac{\sin^2\eta}{2\hbar^2}\, S(t-t') \ ,
\end{equation}
\begin{equation}
\label{Eq:Longitudinal_Sigma} \Sigma_{01 \leftarrow
01}^{\parallel}(t-t')  = -\frac{\cos^2\eta}{\hbar^2}\, S(t-t')\,
e^{\frac{i}{\hbar}\,\Delta E(t-t')} \ .
\end{equation}
Noting that
${(L_{\rm s})}_{01 \leftarrow 01}=i\Delta E/\hbar$ we obtain from
Eq.~(\ref{Eq:Bloch_Redfield_Tensor})
\begin{eqnarray}
\label{Eq:Gamma_0101_perp}
\Gamma_{01 \leftarrow 01}^{\perp} & = & \frac{i\sin^2\eta}{2\hbar^2} \int
\frac{d\omega}{2\pi}\;\frac{S_X(\omega)}{\omega+(\Delta
E/\hbar)-i0} \ ,\\
\Gamma_{01 \leftarrow 01}^{\parallel} & = & \frac{i\cos^2\eta}{\hbar^2} \int
\frac{d\omega}{2\pi}\;\frac{S_X(\omega)}{\omega-i0} \ ,
\end{eqnarray}
and $\Gamma_{01 \leftarrow 01}=\Gamma_{01 \leftarrow 01}^{\perp}+
\Gamma_{01 \leftarrow 01}^{\parallel}$. The real part
of these entries of the Redfield tensor,
${\rm Re}\,\Gamma_{01 \leftarrow 01} =- T_2^{-1}$,
reproduces the dephasing rate (\ref{T2GR}).

\subsection{Energy renormalization}
\label{Sec:Renormalization}
The analysis of the preceding subsection also shows that the qubit's
level splitting is renormalized by the bath. This effect is the
analogue of the  Lamb shift, i.e. the renormalization of the level
splitting in atoms due to the coupling to the
electromagnetic radiation. As can be seen from Eq.~(\ref{Eq:Redfield})
an imaginary part of  $\Gamma_{01 \leftarrow 01}$ gives rise to such a
renormalization. Only the transverse noise contributes to this
renormalization. From Eq.~(\ref{Eq:Gamma_0101_perp}) we obtain
\begin{equation}
\label{Eq:Im_Gamma_0101}
{\rm Im}\,\Gamma_{01 \to 01} =
- \frac{\sin^2\eta}{2\hbar^2}\;{\rm P} \int
\frac{d\omega}{2\pi}\;\frac{S_X(\omega)}{\omega-(\Delta E/\hbar)}
\ .
\end{equation}

 As an  example we consider an oscillator bath with Ohmic
spectral density up to a high-frequency cut-off, $J(\omega) = 2\pi
\alpha\hbar\omega \Theta(\omega_c-\omega)$,
at low temperature $k_{\rm B}T\ll\Delta E$ where $S(\Delta E) =
J(\Delta E)\coth\frac{\Delta E}{2k_{\rm B}T} \approx J(\Delta E)$,
which is coupled purely transversely, $\eta=\pi/2$.
Then the integral in (\ref{Eq:Im_Gamma_0101}) is dominated by high frequencies
$\hbar\omega > \Delta E$ and gives  the following correction to the
energy splitting: $ \Delta E
\rightarrow   \Delta E (1-\alpha \ln(\hbar\omega_c/ \Delta E))$.
This result indicates that the small parameter of the
perturbative expansion is $\alpha\ln(\hbar\omega_c/\Delta E)$,
rather than $\alpha$.
Also for general bath spectra the high-frequency part of the spectrum
typically leads to a renormalized Hamiltonian. The effect can
be calculated perturbatively as shown above or in the frame of a
renormalization-group approach, which accounts for all orders in the small
parameter. It turns out that in such a situation only
the component of the field perpendicular to the direction of the fluctuating
field is affected by the renormalization~\cite{LeggettRMP,WeissBook}.

The effect of low-frequency
fluctuations can be treated adiabatically: They do not induce transitions
between the qubit states, but they change their energy splitting since the
fluctuations of the transverse component of the `magnetic field' $X^\perp$
increase the average magnitude of the field.

Even if the bare Hamiltonian is time-dependent
the contribution of the high-frequency modes
($\hbar\omega\gg\Delta E$) of the environment
can be absorbed in a renormalization of the Hamiltonian.
In contrast, the analysis of the effect of low-frequency modes
 depends on further details.

\section{Applications}\label{Sec:Appl}

\subsection{Ohmic dissipation in real devices}
\label{Sec:Experiment_Ohmic}

For an Ohmic environment we obtain from Eqs.~(\ref{Eq:T1}) and (\ref{T2GR})
\begin{eqnarray}
T_1^{-1}&=&\pi\alpha\;\sin^2\eta\;
\frac{\Delta E}{\hbar}
\coth\displaystyle\frac{\Delta E}{2k_{\rm B}T}
\label{Eq:relaxation}
\; ,
\\
T_2^{-1}&=&
\frac{1}{2}\;T_1^{-1} +
\pi\alpha\;\cos^2\eta\;
\frac{2k_{\rm B}T}{\hbar}
\label{Eq:dephasing}
\; .
\end{eqnarray}
For the charge qubit the coupling constant $\alpha$ is given by
Eq.~(\ref{Eq:alpha}). The typical impedance of the control line
is $R \approx 50~\Omega$. Since
$R_{\rm Q} \approx 6.5~{\rm k}\Omega$ we obtain
$\alpha \approx 10^{-2}(C_{\rm t}/C_{\rm J})^2$.
With the typical values of the capacitances
$C_{\rm J}\approx 10^{-16}{\rm F}$ and
$C_{\rm g}\approx 10^{-18}{\rm F}$
their ratio can be made as small as $10^{-2}$. Thus $\alpha$ can be as
small as $10^{-6}$. At temperature $T\approx 10$~mK and  for
$\Delta E/k_{\rm B} \approx 1$~K, we estimate
the pure dephasing time $T_2^* \approx 100~\mu$s (for $\eta = 0$)
and the relaxation time $T_1 \approx 1~\mu$s (for $\eta = \pi/2$).
In recent experiments~\cite{Saclay_Manipulation_Science,Delft_Rabi}
similar values of $T_1$ have been observed,
however $T_2$ was much shorter. This has been attributed to the
contribution of the $1/f$
noise (i.e. dominated by low-frequencies), which we discuss below.

\subsection{Validity of the Bloch-Redfield approximation
in the Ohmic case}

As mentioned above, after having solved the Bloch equations one should
verify the validity of the Bloch-Redfield approximation
by checking the self-consistency.
Let us consider the example of a weakly coupled Ohmic
environment, $\alpha\ll1$. The  Bloch-Redfield approximation is justified
provided
the integral in Eq.~(\ref{Eq:Bloch_Redfield_Tensor})
converges on a time scale short compared to the appropriate dissipative time
$T_1$ or $T_2$.\\
(i) For the evaluation of $T_1^{-1}\propto S(\Delta E/\hbar)
\sin^2\eta$ consistency
requires that the noise power in the vicinity of $\omega=\Delta
E/\hbar$ varies smoothly on the scale $T_1^{-1}$. For the Ohmic bath this
condition is always satisfied: in the limit $\Delta E\gg k_{\rm B}T$
it reduces
to $\alpha\Delta E\ll\Delta E$ and in the opposite limit to $\alpha T\ll T$.\\
(ii) For the evaluation of $T_2^{-1}=\frac{1}{2}T_1^{-1} + (T_2^*)^{-1} \propto
\frac{1}{2}S(\Delta E)\sin^2\eta+ S(0)\cos^2\eta$ the condition may differ only
if the rate is dominated by the second term, $T_2^*$. Then the condition is that
the noise power varies weakly on the scale of $(T_2^*)^{-1}\sim
\alpha T (k_{\rm B}/\hbar) \ll T (k_{\rm B}/\hbar) $ around
zero frequency, which is always true.

We conclude that for $\alpha \ll 1$ the Bloch-Redfield approximation is
valid.

\subsection{Environment-dominated dynamics}\label{Sec:Zeno}

Consider a Hamiltonian with purely transverse coupling, $\eta =
\pi/2$:
\begin{equation}
H = -\frac{1}{2}\Delta E\;\sigma_z
+\frac{1}{2}X\;\sigma_x +H_{\rm bath}
\ .
\end{equation}
For this model the Bloch-Redfield tensor becomes:
\begin{equation}
\label{Eq:No_RWA}
\frac{d}{dt}
\left(
\begin{array}{c}
\rho_{00}\\
\rho_{11}\\
\rho_{01}\\
\rho_{10}\\
\end{array}
\right)
=
\left(
\begin{array}{cccc}
-\Gamma_{\uparrow} & \Gamma_{\downarrow} & 0 & 0 \\
\Gamma_{\uparrow} & -\Gamma_{\downarrow} & 0 & 0 \\
0 & 0 & i\Delta E -\Gamma_{\varphi} & \Gamma_{\varphi}^{*} \\
0 & 0 & \Gamma_{\varphi} & -i\Delta E - \Gamma_{\varphi}^{*} \\
\end{array}
\right)
\left(
\begin{array}{c}
\rho_{00}\\
\rho_{11}\\
\rho_{01}\\
\rho_{10}\\
\end{array}
\right)
\ .
\end{equation}
Here $\Gamma_{\varphi} \equiv -\Gamma_{01 \leftarrow 01}$. If the two-level
system is
almost degenerate, $\Delta E\ll \hbar \Gamma_\varphi$, the RWA is not valid,
and one can not neglect the off-diagonal entries
(obtained from diagrams similar to those in
Fig.~\ref{Fig:Gamma_11to00}). Notice that the degeneracy condition, $\Delta E\ll
\Gamma_\varphi$, may still be consistent with the validity condition of the
Bloch-Redfield approximation, $1/\tau_{\rm c} \gg \Gamma_\varphi$. For instance,
for an Ohmic bath in the weak-coupling limit, $\alpha\ll 1$, one finds
$1/\tau_{\rm c} \sim (k_{\rm B}/\hbar) T \gg \alpha (k_{\rm B}/\hbar) T \sim
\Gamma_\varphi$ (the latter equality is valid provided $\Delta E\ll k_{\rm B}
T$).

In this regime, $\Delta E\ll \hbar\Gamma_\varphi$, Eq.~(\ref{Eq:No_RWA}) has a
very slowly decaying eigenmode, with $\hat\rho\propto\hat\sigma_x$, with the
relaxation rate
$(\Delta E)^2/2 \Gamma_{\varphi}$ ($\Gamma_\varphi$ is almost real). This
phenomenon is called the quantum Zeno effect~\cite{Misra_Sudarshan}. The
environment ``observes'' $\sigma_x$ and prevents it from
relaxing. The effect has the remarkable property that the stronger the
dephasing,
$\Gamma_{\varphi}$, the slower the relaxation.

In general, in the dissipation-dominated regime it is more
convenient to treat the spin Hamiltonian as a perturbation on top of the
dissipative dynamics.

\subsection{Time-dependent driving: Rabi oscillations}
\label{Sec:Rabi}

For a general time-dependent magnetic field $\vec B(t)$ the description
of the dissipative dynamics is more complicated.
Here we discuss an example for which the Bloch-Redfield approximation does not
apply in the laboratory frame but can be used in the rotating
frame (or, equivalently, in the interaction representation).

Consider a situation with the purely longitudinal coupling,
\begin{equation}
H = -\frac{1}{2}\Delta E\;\sigma_z
-\frac{1}{2}X\;
\sigma_z +H_{\rm bath}
\ ,
\end{equation}
for which Eqs.~(\ref{Eq:T1}) and (\ref{T2GR}) give
$\, T_1^{-1}=0$ and $\, T_2^{-1}=(1/2\hbar^2)\,S_X(\omega= 0)$.
Let us now apply a resonant transverse driving field:
\begin{equation}
H = -\frac{1}{2}\Delta E\;\sigma_z-\frac{1}{2}\Omega_{\rm R}\;
(\cos\omega t\,\sigma_x - \sin\omega t\,\sigma_y)
-\frac{1}{2}X\;
\sigma_z +H_{\rm bath} \, .
\end{equation}
The subscript R indicates that this field can induce the so-called Rabi
oscillations between the levels. As follows from the discussion in
Section~\ref{Sec:Bloch-Redfield}, one can still use the Bloch equations as long
as the driving is weak enough, $\Omega_{\rm R}\tau_{\rm c}\ll 1$. For a
stronger driving these equations may fail. However,
the transformation to the rotating frame [$H\to \tilde H = \hbar\,\dot U U +
UHU^{\dag}$, with $U=\exp\left(-i\omega \sigma_z \, t/2\right)$ and
$\hbar\omega = \Delta E$] makes
the Hamiltonian time-independent:
\begin{equation}
\label{Eq:RabiH}
H= -\frac{1}{2}\Omega_{\rm R}\;\sigma_x
-\frac{1}{2}X\;\sigma_z + H_{\rm bath}
\ .
\end{equation}
The first term on the rhs of Eq.~(\ref{Eq:RabiH}) induces coherent oscillations
in the rotating frame, with the spin performing an oscillatory motion
between the up- and down-states. Furthermore, using Eqs.~(\ref{Eq:T1}) and
(\ref{T2GR}), we find that the relaxation of the $x$-component of the spin in
the rotating frame occurs on the time scale
$T_1^{-1}=(1/2\hbar^2)\,S_X(\omega=\Omega_{\rm R})$, and the decay of the
transverse ($y$- and $z$-) components on the time scale $T_2^{-1}=1/(2T_1)$.
One may verify that in the weak-driving limit ($\Omega_{\rm R}\tau_{\rm c}
\ll 1$) this result in consistent with the standard Bloch equations in the
laboratory frame.

We summarize, strong or fast variations of the Hamiltonian
(e.g., the field $\vec B(t)$) may
invalidate the Bloch equations, but in certain situations one may still reduce
the problem to another one  for which these equation can be applied. The
resulting dissipative times $T_1$, $T_2$ may, however, differ from those in the
original Bloch equations.

\section{Beyond the Bloch-Redfield (golden-rule) approximation}%
\label{Sec:beyond}

\subsection{1/f noise}
\label{Sec:1/f}

Several experiments with Josephson circuits have revealed at low
frequencies the presence of $1/f$ noise. While the origin of this
noise may be different in different circuits, it appears that in
several cases (especially in the charge devices) it derives from
``background charge fluctuations''. This noise is usually
presented as an effective noise of the gate charge (see
Eq.(\ref{Eq:Box_Hamiltonian_n})), i.e., $S_{Q_{\rm g}}(\omega) =
\alpha_{1/f}e^2/|\omega|$. Recent experiments
(cf.~\inlinecite{Nakamura_Echo} and references therein) yield at
relevant temperatures $\alpha_{1/f} \sim 10^{-7}$--$10^{-6}$. We
translate this noise into fluctuations of the effective magnetic
field $X$ with
\begin{equation}
\label{Eq:1/f_spectrum}
S_X(\omega) = \frac{E^2_{1/f}}{|\omega|}
\ ,
\end{equation}
and by comparison with  Eq.~(\ref{Eq:B_z_Q_g}) find $E_{1/f} =
4E_{\rm C}\sqrt{\alpha_{1/f}}$.

The Bloch-Redfield result
(\ref{T2GR}) predicts a divergent rate $T_2^{-1}\to \infty$, unless
$\eta = \pi/2$, and clearly fails in this case. This can be attributed
to the very slow (logarithmic) time decay of the function $S_X(t)$.
It is easy, though, to describe the dephasing due to $1/f$ noise
rigorously, if it couples purely longitudinally ($\eta=0$)
and is Gaussian distributed. Treating
$X$ as a classical variable one obtains~\cite{Cottet_Naples}
\begin{eqnarray}
\label{Eq:Cottet}
&&\langle \sigma_{+}(t)\rangle \propto
\left\langle e^{-\frac{i}{\hbar}\,\int\limits_0^{t}dt'\,X(t')}\right\rangle=
e^{-\frac{1}{2\hbar^2}\int\limits_0^t dt'\;\int\limits_0^t dt'' \langle
X(t')X(t'') \rangle}
\nonumber \\
&&=\exp\left({-\frac{1}{2\hbar^2}\,\int
\frac{d\omega}{2\pi}\,S_X(\omega)\,\frac{\sin^2(\omega t/2)}
{\left(\omega/2\right)^2}}\right)
\ .
\end{eqnarray}
This result coincides with the real part of
Eq.~(\ref{Eq:P(t)_F2}) obtained in a quantum description.
For a regular low-frequency spectrum one may replace the last fraction
in (\ref{Eq:Cottet}) by
$2\pi\delta(\omega)\,t\, $ and obtain the golden-rule result for the dephasing
time. For the $1/f$ noise one obtains with logarithmic
accuracy~\cite{Cottet_Naples}
\begin{equation}
\langle \sigma_{+}(t)\rangle \propto
\exp\left(-\frac{E_{1/f}^2}{2\pi\hbar^2}\,t^2\, |\ln t\omega_{\rm ir}|\right)\
,
\end{equation}
where $\omega_{\rm ir}$ is the infrared cutoff frequency for the $1/f$
noise.~\footnote{In an experiment with averaging over repetitive
measurements the infrared cutoff may be determined by the time
interval $t_{\rm av}$ over which the averaging is
performed~\cite{Saclay_Manipulation_Science}. In a
spin-echo experiment it may be set by the echo
frequency~\cite{Nakamura_Echo}.} From this decay law one can deduce
the dephasing time scale, \begin{equation}
\label{Eq:T2_1/f}
\frac{1}{T_2^*} \approx
\frac{1}{\hbar}\,
E_{1/f}\, \sqrt{\frac{1}{2\pi} \ln \frac{E_{1/f}}{\omega_{\rm ir}}}
\ .
\end{equation}
Using experimental parameters for
an estimate one obtains $T_2$ in the range from fractions to a
few nanoseconds, which is close
to the dephasing time observed in the experiments away from the special
point $\eta =\pi/2$. We also see that this source of dephasing
dominates over the Ohmic noise.

\subsection{Contribution of higher orders}
\label{Sec:2order}

In this section we consider a further situation where the lowest-order Bloch
equations are not
sufficient. The dissipative times (\ref{Eq:T1}, \ref{T2GR}) in the Bloch
equations are dominated by the noise power at specific frequencies ($0$ and
$\Delta E/\hbar$).
 As an example, consider the purely transverse case, $\eta=\pi/2$,
 where the first-order Bloch-Redfield formulas give $T_1^{-1}=2
 T_2^{-1}=\frac{1}{2\hbar^2}S_X(\omega=\frac{\Delta E}{\hbar})$.
In higher orders of the perturbative expansion
(cf.~Section~\ref{Sec:Diag}) the noise power in other frequency ranges may be
involved. If the noise power in these ranges is much stronger,
higher-order contributions may compete with the lowest-order.
An important example is the case of $1/f$ noise. Here the strong noise at low
frequencies contributes in higher order to the dissipative rates.

\begin{figure}
\centerline{\hbox{\psfig{figure=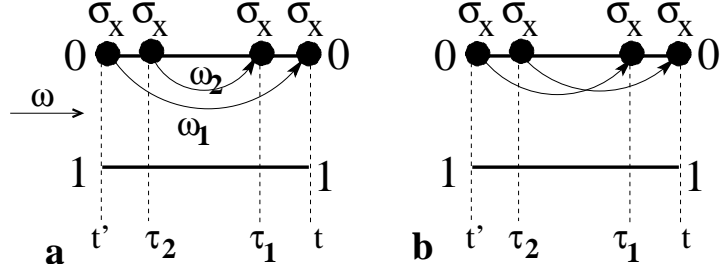,%
width=0.8\columnwidth}}}
\caption[]{Second-order contributions. One can obtain all 16 second-order
diagrams by shifting an even number of the vertices in {\bf a} or {\bf b} to the
lower line.}
\label{Fig:2order}
\end{figure}

To illustrate the importance of higher orders, consider the contribution to the
second-order self-energy $\Sigma_{01 \leftarrow 01}^{(2)}$ shown in
Fig.~\ref{Fig:2order}a,
\begin{equation}
\label{Eq:2order_time}
\delta \Sigma_{01 \leftarrow 01}^{(2)} =
\frac{1}{(2\hbar)^4} \int\limits_{t'}^{t} d\tau_1
\int\limits_{t'}^{\tau_1} d\tau_2\,
\langle X(t)X(t')\rangle\,\langle X(\tau_1)X(\tau_2)\rangle
e^{i\Delta E(\tau_1 - \tau_2)/\hbar}
\ .
\end{equation}
The factor $e^{\frac{i\Delta E}{\hbar}(\tau_1 - \tau_2)}$ is due to the
difference in energy of the state on the upper ($\ket{0}$) and lower ($\ket{1}$)
lines in the time interval between $\tau_1$ to $\tau_2$.
The Fourier transform of Eq.~(\ref{Eq:2order_time}) reads
\begin{eqnarray}
\label{Eq:2order_Laplace}
\delta \Sigma_{01 \leftarrow 01}^{(2)}(\omega) = &&
\frac{-i}{(2\hbar)^4} \int\limits_{-\infty}^{\infty}
\frac{d \omega_1 d \omega_2}{(2\pi)^2}\,
\langle X^2_{\omega_1}\rangle \langle X^2_{\omega_2}\rangle \nonumber\\
&& \times \frac{1}
{(\omega-\omega_1+i0)^2
(\omega-\omega_1-\omega_2+\Delta E/\hbar+i0)}
\ .
\end{eqnarray}
Comparison of Eq.~(\ref{Eq:2order_Laplace}) and
Fig.~\ref{Fig:2order}a elucidates the diagrammatic
rules: the bath lines carry the frequencies $\omega_1$
and $\omega_2$, while pairs of horizontal solid lines contribute factors
of the type $1/(\omega-\sum \omega_n-i\hat L_{\rm s})$, where the summation is
over all the bath lines in the interval and the matrix
element of $L_{\rm s}$ is determined by the spin states on the upper and lower
lines.

There are in total 16 second-order diagrams (see
Fig.~\ref{Fig:2order}), which contribute to $\Sigma_{01 \leftarrow 01}$.
They add up to
\begin{eqnarray}
\label{Eq:2order_All}
\Sigma_{01 \leftarrow 01}^{(2)}(\omega) =&&
\frac{-2i}{(2\hbar)^4} \int\limits_{-\infty}^{\infty}
\frac{d \omega_1 d \omega_2}{(2\pi)^2}\,
\left[\frac{1}{\omega-\omega_1+i0} + \frac{1}{\omega-\omega_2+i0}\right]
\nonumber \\
&&
\times \frac{1}{\omega-\omega_1+i0}
\left[\frac{\langle X^2_{\omega_1} \rangle \langle X^2_{\omega_2}\rangle +
\langle X^2_{-\omega_1} \rangle \langle X^2_{-\omega_2}\rangle}{%
\omega-\omega_1-\omega_2+\Delta E/\hbar+i0} \right.\nonumber \\
&&
\quad \quad \quad \quad
+ \left. \frac{\langle X^2_{\omega_1} \rangle\langle X^2_{-\omega_2}\rangle+
\langle X^2_{-\omega_1}\rangle \langle X^2_{\omega_2}\rangle}{%
\omega-\omega_1-\omega_2-\Delta E/\hbar+i0}\right]
\ .
\end{eqnarray}
We evaluate the behavior of the self-energy in the
vicinity of the level splitting, $\Sigma_{01
\leftarrow 01}^{(2)}(\omega=-\Delta E/\hbar -\omega' +i0)$, where $\omega' \ll
\Delta
E/\hbar$. If the integral (\ref{Eq:2order_All}) is dominated by low frequencies,
we obtain:
\begin{eqnarray}
\label{Eq:2order_BR}
{\rm Re}\,\Sigma_{01 \leftarrow 01}^{(2)} \Bigl( \omega&&=-\frac{\Delta
E}{\hbar} - \omega' +i0 \Bigr)
\nonumber\\
&& \approx -
\frac{1}{8\hbar^2\Delta E^2} \int\limits_{-\infty}^{\infty}
\frac{d \nu}{2\pi}\,
\left[\langle X^2_{\nu+\omega'}\rangle \langle X^2_{-\nu}\rangle +
\langle X^2_{\nu-\omega'}\rangle \langle X^2_{-\nu}\rangle\right]
\nonumber\\
&&=-\frac{1}{2\hbar^2} S_{X_2}(\omega')
\ ,
\end{eqnarray}
where $S_{X_2}$ is the symmetrized correlator of $X_2 \equiv X^2/(2\Delta E)$.

To understand the origin of this result (\ref{Eq:2order_BR}) consider the effect
of the low-frequency fluctuations (which, we assumed, dominate the dephasing).
Starting from the Hamiltonian (\ref{Eq:Spin_Boson_Eigen_Basis}) with
the transverse coupling to the noise source, $\eta=\pi/2$, we see that in the
adiabatic limit of very slow fluctuations the dynamics reduces to an
accumulation of the relative phase between the eigenstates, $\int dt\,\Delta
E_X(t)/\hbar$, where $\Delta E(X) = \sqrt{\Delta E^2 + X^2} \approx \Delta E +
X^2/(2\Delta E)$. Thus the problem reduces to the longitudinal case. The
calculation above confirms this reduction in the lowest order. More precisely,
the second-order contribution to $\Sigma_{01 \leftarrow 01}^{(2)}$ for the
transversely coupled $X$ coincides with the lowest-order contribution for the
longitudinally coupled $X_2$.

Notice also that the fluctuating field $X_2$ has a nonzero average which
provides a (low-frequency contribution to the) renormalization of the level
splitting (cf.~Section~\ref{Sec:Renormalization}).

Generally speaking, the result (\ref{Eq:2order_BR}) does not allow one to
evaluate the dephasing decay law but may be used for qualitative estimates.
Consider the example of the $1/f$ noise. In this case the low-frequency behavior
of the second-order self-energy is singular and it may compete with the
first-order contribution. Higher-order contributions require further analysis
(both in the linear transverse and quadratic longitudinal cases). However, one
can show that at high enough frequencies the higher-order contributions to
$\Sigma$ may be neglected. Thus, the result (\ref{Eq:2order_BR}) permits an
analysis of the decay law at short times. Extrapolating this law to longer
times,
one can estimate the typical dephasing time. Qualitatively, this amounts to
solving the equation $T_2^{-1} \sim -\Sigma_{01 \leftarrow 01}^{(2)}
(\omega=T_2^{-1})$. For $1/f$ spectrum (\ref{Eq:1/f_spectrum}) we obtain
\begin{eqnarray}
\label{Eq:SX21/f}
S_{X_2}(\omega)
\sim \frac{E_{1/f}^4}{(\Delta E)^2}\,\frac{1}{|\omega|}\ln\frac{|\omega|}
{\omega_{\rm ir}}
\ ,
\end{eqnarray}
i.e., this noise power is again of the $1/f$ type.
This gives the estimate $T_2^{-1} \sim (E_{1/f}^2/\hbar\Delta
E) \ln(E_{1/f}^2/\hbar\Delta E\omega_{\rm ir})$.




\subsection{Effect of strong pulses}
\label{Sec:Pulse}

In this section we discuss the
effects of a strong pulse of the magnetic field on the dissipative
dynamics of the spin.
The derivation of the Bloch equations
(cf.~the discussion in Section~\ref{Sec:Bloch-Redfield})
relied on the slowness of the spin's dynamics, which may be violated
if a strong or a fast-oscillating field is applied.
In this case  for a period of order $\tau_{\rm c}$
a different description beyond the Bloch-Redfield approximation is
needed.

As an example we consider the `initial-state preparation' of a gedanken
experiment designed to observe the effects of dephasing on the spin precession.
Specifically, we assume that initially the spin was kept at low temperatures in
a field $B_z\hat z$ and has relaxed to the ground state $\ket{\uparrow}$.  Then
it is rotated by a strong short $\pi/2$-pulse of the transverse field and starts
a free precession in the field $B_z\hat z$.  Under the influence of the bath the
relative phase between the spin-up and -down states will not only have a
contribution due to the Zeeman splitting, which grows linearly in time, but also
a random contribution from the noise, which destroys the phase coherence. For
the analysis below we assume the Ohmic spectral density.

The effect of the pulse on the spin's dynamics, as compared to the
Bloch equations, is an additional unitary evolution and dephasing.
The former effect is an extra phase acquired by the spin due to the initial
polarization of the bath. The latter is the dephasing due to the initial partial
disentanglement of the bath and the spin (which amounts to a partial
factorization of the initial density matrix), as we illustrate below.

During the relaxation of the spin to the  ground state
$\ket{\uparrow}$ the oscillators
have also relaxed to the corresponding ground state. I.e.,
the initial state of the whole system is given by the density
matrix $\rho_0 = \ket{\uparrow}\bra{\uparrow}\otimes
\rho_{\uparrow}$ with $\rho_{\uparrow} \propto
\exp(-H_{\uparrow}/k_{\rm B}T)$ and $H_{\uparrow}\equiv H_{\rm bath} - X/2$.
In this state the bath operator $X$ provides a finite average
field. Its expectation value
$\langle X \rangle$ can be calculated as a linear response of the bath to
the perturbation $H_{\rm pert}=-fX$ with $f=1/2$, i.e.,
$\langle X\rangle = \chi(\omega=0) f = \chi'(\omega=0)/2$.
For an Ohmic spectrum, $\chi''(\omega) = 2\pi\alpha\hbar\omega$ up to a cutoff
$\omega_{\rm c}$, we have $\langle X\rangle =
\chi'(\omega=0)/2 = \int (d\nu\,\chi''(\nu))/(2\pi\nu) \sim \alpha
\hbar \omega_{\rm c}$. We note that this expectation value depends  strongly
on the high-frequency cutoff $\omega_{\rm c}$ of the bath spectrum.
The field produced by each bath mode averages
out after its typical oscillation period; still during a short initial period of
length $\sim\omega^{-1}$ it affects the spin dynamics. Depending on
the spectral density of the bath, the resulting effect may be
dominated by low or high frequencies.

Now we apply a $\pi/2$-pulse to rotate the spin to a position
perpendicular to the original state, i.e. into the superposition
$(\ket{\uparrow} + \ket{\downarrow})/\sqrt{2}$. If the
pulse duration is very short, the state of the bath does not change during the
pulse. If the pulse takes a finite time, $\omega_p^{-1}$, the bath oscillators
partially adjust to the changing spin state. For the high-frequency oscillators,
$\omega \gg \omega_p$, the spin evolution during the pulse (and afterwards) is
adiabatic; they just `dress' the spin, renormalizing its parameters, e.g., its
$g$-factor~\cite{Shnirman_Makhlin_Schoen_Nobel}. In contrast, the  slow
oscillators, with  $\omega\ll\omega_p$, do not have time to
react on this short time interval and start to change their states only after
the pulse. They react differently for the spin-up and spin-down components, thus
contributing to the dephasing. Neglecting, for a qualitative discussion,
the oscillators at intermediate frequencies, we find the system in the state
$\rho_0 = \frac{1}{2}\ket{\tilde\uparrow+\tilde\downarrow}
\bra{\tilde\uparrow+\tilde\downarrow}\otimes \tilde\rho_{\uparrow}$.
Here the
state $\ket{\tilde\uparrow}$ denotes the state of the dressed spin, i.e.
the product state of $\ket{\uparrow}$ and the equilibrium states of high
frequency oscillators under the influence of the spin up (i.e. with the
Hamiltonian $H_\uparrow$), and similarly for the state $\ket{\tilde\downarrow}$.
In short, qualitatively one may think of the system as a dressed spin coupled to
a bath with the upper cutoff $\sim\omega_p$.

After this initial preparation, the spin starts a precession in the field
$B_z \hat z$.
To describe the time evolution in the presence of the bath
we split the Hamiltonian (\ref{Eq:Spin_Boson_Hamiltonian}) as
\begin{equation}
\label{Eq:Spin_Boson_Rewritten} H=-\frac{1}{2} B_z\sigma_z -
\frac{1}{2} X(\sigma_z-1) + H_{\uparrow}
\end{equation}
and extract explicitly the average $\langle X \rangle$ with the
intention to treat the part proportional to $\delta X$ as a
perturbation. Hence
\begin{equation}
\label{Eq:Spin_Boson_Shifted} H=-\frac{1}{2} B_z\sigma_z -
\frac{1}{2} \langle X\rangle (\sigma_z-1)
-\frac{1}{2} \delta X (\sigma_z-1) +H_{\uparrow}\  .
\end{equation}

At $t = 0$ the spin is subject to the additional
field $\langle X \rangle$. We expect that this field relaxes, on the time scale
$\omega^{-1}$ for oscillators with frequency $\omega$. Nevertheless, before the
field vanishes, it contributes to the phase evolution.
Clearly, the spin-up and spin-down states are decoupled, which allows for an
exact solution. One may use the so called polaron transformation
(cf.~\inlinecite{LeggettRMP}, \inlinecite{WeissBook}), or
calculate the time
evolution of the off-diagonal element of the qubit's density
matrix directly:
\begin{equation}
\label{Eq:Sigma_plus_Time_Evolution}
\langle\sigma_+(t)\rangle =  e^{-\frac{i}{\hbar}B_z t}\;
{\rm Tr}\,\left[ \sigma_+\; S(t,0)\; \rho_0\; S^{\dag}(t,0)
\right]
\ ,
\end{equation}
where
\begin{equation}
S(t,0) \equiv T e^{\frac{i}{2\hbar}\int\limits_{0}^{t}
\delta X(t')(\sigma_z-1)\,dt'}
\ .
\end{equation}
Obviously, the evolution operators $S$ and $S^{\dag}$
do not flip the spin, while the operator $\sigma_+$ in
Eq.~(\ref{Eq:Sigma_plus_Time_Evolution}) imposes
a selection rule such that the spin is in the
state $\ket{\downarrow}$ on the forward Keldysh contour (i.e., in
$S(t,0)$) and in the state $\ket{\uparrow}$ on the
backward contour (i.e., in $S^{\dag}(t,0)$).
Since $\bra{\uparrow}\sigma_z - 1\ket{\uparrow} = 0$ the
contribution of the backward contour vanishes.
From $\bra{\downarrow}\sigma_z - 1\ket{\downarrow} = -2$
we obtain
$\langle\sigma_+(t)\rangle = P(t)\exp(-iB_z
t/\hbar)\langle\sigma_+(0)\rangle$,
where
\begin{equation}
\label{Eq:P(t)_from_up} P(t)\equiv \langle T
\exp\left(-\frac{i}{\hbar}\int\limits_{0}^{t} \delta X(t')
dt'\right) \rangle \ .
\nonumber \\
\end{equation}
Due to the harmonic nature of the bath
the equilibrium correlation functions of $\delta X$ for a bath with the
Hamiltonian $H_{\uparrow}$ coincide with those for $X$ and $H_{\rm bath}$.
Since the fluctuations of $\delta X$ are Gaussian, we find
\begin{eqnarray}
\label{Eq:P(t)_F2} \ln P(t) &=&
-\frac{i}{2\hbar^2}\int\limits_{0}^{t}\int\limits_{0}^{t}
d t_1 d t_2 G^{c}(t_1-t_2)\nonumber \\
&=& -\frac{i}{2\hbar^2}
\int\,\frac{d\omega}{2\pi}\,G^{c}(\omega)\,\frac{\sin^2(\omega
t/2)} {(\omega/2)^2} \ .
\end{eqnarray}
In equilibrium the Green function $G^{c}(t-t') = -i\langle T \delta X(t)\delta
X(t')\rangle$ is related to the noise by the fluctuation-dissipation theorem:
\begin{equation}
G^{c}(\omega) = -\hbar\left(\chi'(\omega) +
i\chi''(\omega)\coth\frac{\omega}{2T}\right)= -\hbar\chi'(\omega)
- i S(\omega) \ .
\end{equation}

If we substituted $\sin^2(\omega t/2)/(\omega/2)^2 \to
2\pi\delta(\omega)\,t$ in
Eq.~(\ref{Eq:P(t)_F2}) we would obtain
\begin{equation}
\label{Eq:P(t)_Golden_Rule} \ln P(t) = -\frac{1}{2\hbar^2}
S_X(\omega=0)\;t + \frac{i}{2\hbar}\chi'(\omega=0)\;t \ .
\end{equation}
The first term in Eq.~(\ref{Eq:P(t)_Golden_Rule}) yields the standard
longitudinal dephasing rate $1/T_2^*=2\pi\alpha k_{\rm B}T/\hbar$,
while the second term exactly cancels the effect of the additional magnetic
field $\langle X \rangle$. Thus, this approximation
predicts a vanishing dephasing rate at $T=0$, and there is no net effect of the
initial preparation.

However, the substitution above is justified only if
$G^c(\omega)$ has no structure at low frequencies $\omega\sim1/t$. This
condition is violated, for an Ohmic bath, for not too long times, still small
compared to
the inverse temperature $\omega_p^{-1} <t<\hbar/k_{\rm B}T$ (cf.~e.g.,
\inlinecite{Shnirman_Makhlin_Schoen_Nobel}). In this time range  we
find ${\rm Re}\,\left[\ln
P(t)\right] \approx -2\alpha|\ln(\omega_{\rm p} t)|$,
which implies a power-law decay
\begin{equation}
\label{Eq:power_law_dephasing}
|\langle\sigma_+(t)\rangle| = (\omega_{\rm p} t)^{-2\alpha}
|\langle\sigma_+(0)\rangle|
\ .
\end{equation}
Thus even at $T=0$, when $S_X(\omega=0)=0$, the off-diagonal elements
of the density matrix decay in time.
The reason is the excitation of the low-frequency ($\omega<\omega_p$) modes of
the environment by the initial strong pulse. Only the excited modes
contribute to the dephasing. From this viewpoint, the effect of the initial
pulse is similar to heating.

Finally, the extra phase contribution
$\delta \Phi(t) \equiv {\rm Im}\,\left[\ln P(t)\right] -
\langle X \rangle\,t/\hbar$ is
\begin{eqnarray}
\label{Eq:Extra_Phase} \delta \Phi &=&
\frac{1}{2\hbar}
\int\,\frac{d\omega}{2\pi}\,\chi'(\omega)\,
\left[\frac{\sin^2(\omega t/2)}{(\omega/2)^2}
-2\pi\delta(\omega)\,t\right]
\nonumber \\
&=& - \frac{1}{\hbar}\int\,\frac{d\omega}{2\pi}\,
\frac{\chi''(\omega)}{\omega^2}\, \sin\omega t \ .
\end{eqnarray}
In deriving Eq.~(\ref{Eq:Extra_Phase}) we have used the analyticity
of $\chi$ in the upper half-plane. For the
Ohmic bath, $\delta \Phi
\rightarrow -\pi\alpha$ for $t\gg \omega_{\rm p}^{-1}$. Thus the extra
phase is acquired during a short period of time of order
$\omega_{\rm p}^{-1}$. One may attribute this extra phase to the additional
magnetic field $\langle X \rangle$ produced by the
excited modes in the bath.
Note also that the time, over which the extra phase is acquired, may depend on
the power spectrum of the noise.

After a period of free precession a measurement of the
final spin state is to be performed in order to monitor the effect of
dissipation. A `low-energy measurement',
which does not break the adiabaticity condition for the high-frequency
modes, reads out the state of the dressed spin.

\subsection{`Optimal' operation points and nonlinear coupling}
\label{Sec:X2}

In the presence of $1/f$ noise, its longitudinal component typically
dominates dephasing (see Section~\ref{Sec:1/f}).
By adjusting the system control parameters to a point where the longitudinal
 coupling to  the low-frequency noise is tuned to zero one should be able
to increase the coherence time. This increase was observed, indeed, in
recent experiments~\cite{Saclay_Manipulation_Science}. The qubit in
this experiment was subject to several electromagnetic noise
 sources, including charge and flux noise. When tuned to the optimal
 operation point  the remaining coupling to
 the noise is linear transverse for one part, $X$, of the noise
 sources  and quadratic longitudinal for the rest, $Y$.
In the mentioned experiments $X$
refers to the charge noise, while $Y$ describes flux noise. The
resulting Hamiltonian reads
\begin{equation}
\label{Eq:Vion}
H = -\frac{1}{2} \left(
\Delta E\,\sigma_z + X\, \sigma_x + Y_2\, \sigma_z
\right) \, .
\end{equation}
Here, similar to the discussion after Eq.~(\ref{Eq:2order_BR}),
we introduced $Y_2\equiv Y^2/(2E_Y)$. The energy scale
$E_Y$ characterizes the strength of the quadratic coupling
to the noise source $Y$. For the setup of
\inlinecite{Saclay_Manipulation_Science}
$E_Y$ is of the order of the qubit's Josephson energy.

For noise spectra with a low-frequency singularity (e.g., $1/f$), both the
transverse linear and the longitudinal quadratic coupling can induce  strong
dephasing. As discussed in
Section~\ref{Sec:2order}, for a transverse coupling to a source of
low-frequency noise it may be necessary to
account for higher-order contributions in the calculation of
the dissipative rates.
In particular, we have seen that, at least in the second order, the
contribution of the transverse noise $X$ reduces to that of
the longitudinally coupled field $X_2 = X^2 /(2\Delta E)$.
Thus, both the linear transverse coupling and the quadratic
longitudinal coupling in (\ref{Eq:Vion}) have similar effects on the
dephasing. In this approximation, the effective low-frequency Hamiltonian reads
\begin{equation}
\label{Eq:Vion_effective}
H = -\frac{1}{2}
\Delta E\,\sigma_z  -\frac{1}{2}(X_2 + Y_2)\, \sigma_z\, .
\end{equation}
We neglected the correlations between $X$ and $Y$, and can estimate their
effects independently: $\langle \sigma_{+}(t)\rangle \propto
P_{X_2}(t)P_{Y_2}(t)$,
where the functions $P_{X_2}$ and $P_{Y_2}$ characterize the dephasing
due to the respective noise sources.
As an example we estimate $P_{X_2}(t)$.
The noise power of the variable $X_2$ is given in Eq.~(\ref{Eq:SX21/f}).
Treating $X_2$ as Gaussian distributed
we obtain by substituting Eq.~(\ref{Eq:SX21/f}) in
Eq.~(\ref{Eq:Cottet}) the decay law  (see Table~\ref{Tab:sum})
\begin{equation}
\label{Eq:P_X}
P_{X_2}(t)\rangle \approx
\exp\left(-\frac{E_{1/f}^4}{\hbar^2\Delta E^2}\, t^2 \ln^2(\omega_{\rm ir}t)
\right) \,
\end{equation}
and an estimate for the dephasing rate,  $1/T_2^* \sim (E_{1/f}^2/\hbar
\Delta E) \ln[E_{1/f}^2/\Delta E\omega_{\rm ir}]$.
Note, however, that the result (\ref{Eq:P_X})
is valid only at short times $t\ll T_2^*$. At longer times $t\sim T_2^*$
the non-Gaussian nature of the fluctuations of $X_2$ becomes
important~\cite{Our_Flicker}.
These, however, are not expected to substantially modify the estimate for the
dephasing time scale obtained from an extrapolation of (\ref{Eq:P_X}).
In conclusion, we note that for $\Delta E \gg E_{1/f}$
the dephasing due to $1/f$ noise at the optimal point is substantially
reduced as compared to the general case where $X$ is coupled linearly
to $\sigma_z$.


\begin{table}
\begin{tabular}{|l|l|l|l|l|}
\hline
& Bloch-Redfield & $1/f$, $\parallel$ &
$X^2$, $1/f$, $\parallel$ &Ohmic, $\parallel$

\\
&&$S(\omega)\propto E_{1/f}^2/\omega$&
$S_{X_2}(\omega)\propto \ln\frac{|\omega|}{\omega_{\rm ir}}/\omega$ &
$S(\omega)\propto \alpha\omega$
\\
&\null\hfill Sections~\ref{Sec:Diss}, \ref{Sec:Appl}
&\null\hfill Section~\ref{Sec:1/f}
&\null\hfill Section~\ref{Sec:X2} &\null\hfill Section~\ref{Sec:Pulse}
\\
\hline
$\rho_{01}\propto$& $e^{-t/T_2}$& $e^{-E_{1/f}^2t^2|\ln(\omega_{\rm ir}t)|}$&
$e^{-\frac{E_{1/f}^4}{\Delta E^2} t^2 \ln^2(\omega_{\rm ir}t) }$&
$(\omega t)^{-2\alpha}$ for $t<\frac{\hbar}{k_{\rm B}T}$\\
\hline
\end{tabular}
\caption{The decay law for the off-diagonal (in the qubit's
eigenbasis) entries of the
qubit's density matrix.
The Bloch-Redfield approximation predicts an exponential decay
with dephasing time $\propto S^{-1}(\omega=0)$.
Three examples are listed where this time vanishes or diverges
and a more careful analysis is required.
The sign $\parallel$ indicates the longitudinal
coupling ($\eta=0$).}
\label{Tab:sum}
\end{table}

\section{Summary}\label{Sec:Summary}

In this tutorial we have discussed the issue of dissipation
in solid-state qubits due to the coupling
to the environment. First, the electromagnetic environment
of the charge qubit was described and the coupling Hamiltonian was
derived, and the relevant Caldeira-Leggett model was introduced.
For weak dissipation  the description
by the Bloch equations is sufficient in many situations. We have presented the
calculation of the relaxation and dephasing times, $T_1$ and $T_2$,
in the Bloch-Redfield approximation
and analyzed their validity range. We have also considered examples
where the Bloch-equation description is not sufficient. This includes the
effects of the strong pulses and dephasing due to $1/f$ noise.
Some of the examples analyzed are illustrated in Table~\ref{Tab:sum}. It
includes
situations in which the Bloch-Redfield approximation is insufficient
since it would produce vanishing or diverging
dephasing rates.\\

The work is part of the EU IST
Project SQUBIT and of the {\bf CFN} (Center for Functional Nanostructures)
which is supported by the DFG (German Science Foundation).
Y.M. was supported by the Humboldt foundation, the BMBF, and the ZIP
Programme of the German government.

\newpage

\section{Appendix: derivation of the Hamiltonian}\label{Sec:App}
In this section we present a way of
introducing the environment following \inlinecite{Ingold_Nazarov},
in which the bath is modeled by a transmission line. Even if
a real device is not coupled to an $LC$-line or a resistor,
in generic situations the effect of the
fluctuations on the system (qubit) can be {\it modeled} by an
$LC$-line. The transmission line is also equivalent to an oscillator
bath model introduced by Caldeira and Leggett. The purpose of the
following is to present  a particular  model
which might clarify  some ideas behind the Caldeira-Leggett
approach. A similar analysis for a slightly different model was presented by
\inlinecite{Paladino_LCline}.
\begin{figure}
\centerline{\hbox{\psfig{figure=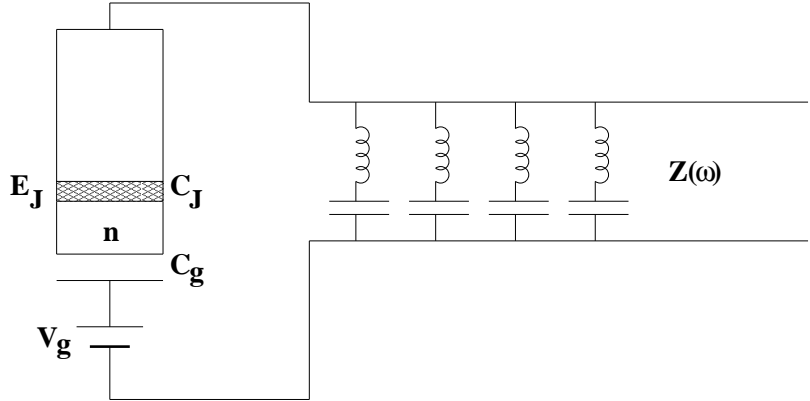,width=
0.9\columnwidth}}}
\caption[]{The qubit's environment modeled by a transmission line.}
\label{Fig:Alternative_Qubit_Environment}
\end{figure}

Consider a transmission line, or $LC$-line, shown in
Fig.~\ref{Fig:Alternative_Qubit_Environment}. It amounts to an impedance
$Z(\omega)$
\begin{equation}
Z^{-1}(\omega) =
\sum_a \left( -i\omega L_a - {1\over i\omega C_a}+0 \right)^{-1}
=
\sum_a {i\omega \over
L_a (\omega^2 - \omega^2_a +i0\mathop{\rm sign}\omega)}
\ ,
\end{equation}
with $\omega_a \equiv 1/\sqrt{L_a C_a}$ (index $a$ counts the
$LC$ elements of the transmission line). The infinitesimal resistance in each
$LC$-element serves to regularize the model. One way
to derive the Hamiltonian is to start by introducing the degrees of freedom
(the phase drops at circuit elements), writing the Lagrangian of the system (a
combination of the charging and Josephson energies), and then performing the
Legendre transformation. For convenience instead of the phase variables
$\theta_k$ ($k$ being any index) we use the dimensionful flux variables
$\phi_k=\Phi_0\theta_k/2\pi$
(defined as time integrals of respective voltage drops; here $\Phi_0=h/2e$), to
which we still refer as `phases'. As independent variables one can use
the phase on the Josephson junction $\phi_{\rm J}$, the phase
between the terminals of the impedance $\phi$ and the phases on the
capacitors of the $LC$ elements of the transmission line $\phi_a$. Then the
phases on the inductances of the $LC$ elements are given by $\phi - \phi_a$,
while the phase on the gate capacitor is given by the Kirchhof rule
$\phi_{\rm g}=-(\phi_{\rm J} + \phi + \int V_{\rm g} dt')$.
The Lagrangian of the system reads
\begin{equation}
L= \frac{C_{\rm J}\,\dot\phi_{\rm J}^2}{2}+
\frac{C_{\rm g}\,(\dot\phi_{\rm J}+\dot\phi+V_{\rm g})^2}{2}
+ E_{\rm J}\cos\left(2\pi\frac{\phi_{\rm J}}{\Phi_0}\right)
+L_Z(\phi)
\ ,
\end{equation}
where
\begin{equation}
L_Z(\phi) \equiv \sum_a \left[
\frac{C_a \dot\phi_a^2}{2} - \frac{(\phi_a - \phi)^2}{2 L_a}
\right]
\ .
\end{equation}
From here it is straightforward to derive the Hamiltonian of the system,
\begin{equation}
\label{Eq:Charge_Box_Hz}
H =
\frac{(2en-q)^2}{2C_{\rm J}}
-E_{\rm J} \cos \theta + \frac{(q-C_{\rm g}V_{\rm g})^2}{2C_{\rm g}} +
H_Z(\phi)
\ ,
\end{equation}
where
\begin{equation}
H_Z(\phi) \equiv \sum_a \left[ \frac{q_a^2}{2C_a} + \frac{(\phi_a
- \phi)^2}{2 L_a} \right] \ .
\label{Eq:HZ}
\end{equation}
Here $(n,\theta)$, $(q,\phi)$, $(q_a,\phi_a)$ are the pairs of conjugate
variables, and $q=-C_{\rm g}\dot\phi_g$ is the charge on the gate capacitor. We
assumed that the charge of the island $2en$ is completely located
on the plates of
the capacitors $C_{\rm J}$ and $C_{\rm g}$, and neglected the so-called offset
charge induced by parasitic voltages due, e.g., to extra charges in the
substrate. Introducing the new charge $\tilde q \equiv q-C_{\rm t}V_{\rm g}$ we
obtain
\begin{equation}
\label{Eq:Hamiltonian_Vg_direct}
H =
\frac{(2en-C_{\rm t}V_{\rm g})^2}{2C_{\rm J}}
-E_{\rm J} \cos \theta - \frac{2e}{C_{\rm J}} n\tilde q
+ C_{\rm t}\dot V_{\rm g} \phi +
\frac{\tilde q^2}{2C_{\rm t}} + H_Z(\phi)
\,,
\end{equation}
thus recovering Eq.~(\ref{Eq:Hamiltonian_CL}).

Indeed, the degrees of freedom
$q_a,\phi_a$ and $q,\phi$ play the part of the oscillator
bath, its Hamiltonian being given by the last two terms in
Eq.~(\ref{Eq:Hamiltonian_Vg_direct}) (its eigenmodes are not given explicitly).
Further, $\tilde q/C_{\rm t}=\delta V$ corresponds to the voltage fluctuations,
and the first term in Eq.~(\ref{Eq:Hamiltonian_Vg_direct}) coincides with the
sum of the first and the last terms in Eq.~(\ref{Eq:Hamiltonian_CL}).
Notice that the last two terms in Eq.~(\ref{Eq:Hamiltonian_Vg_direct})
describe a parallel connection of the impedance and an effective capacitance
$C_{\rm t}$. Accordingly -- one can check this also by an explicit
calculation --
the produced noise is  given by Eq.~(\ref{Eq:S_V}), i.e., the bath
`corresponds' to the impedance $Z_{\rm t}$ rather than $Z$.

In this model one can also observe the result of an abrupt change of the gate
voltage. It is most easily seen from Eqs.~(\ref{Eq:Charge_Box_Hz},
\ref{Eq:HZ}) that
the gate voltage is coupled to the qubit's degree of freedom $n$ only via
the damped mode $(q,\phi)$. As a result, for a pure resistor
$Z(\omega) = R$ changes of
$V_{\rm g}$ influence the qubit only after a delay time,
of order $RC_{\rm t}$. In the form (\ref{Eq:Hamiltonian_Vg_direct})
the term proportional to the time derivative of $V_{\rm g}$ is
responsible for this delay.


\bibliographystyle{klunamed}
\bibliography{ref}
\end{article}
\end{document}